\documentclass{article}
\usepackage{ijcai24}

\usepackage{times}
\usepackage{soul}
\usepackage{url}
\usepackage[hidelinks]{hyperref}
\usepackage[utf8]{inputenc}
\usepackage[small]{caption}
\usepackage{graphicx}
\usepackage{amsmath}
\usepackage{amsthm}
\usepackage{mathtools}
\usepackage{amssymb}
\usepackage{booktabs}
\usepackage{comment}
\usepackage{tabularray}
\usepackage{stfloats}
\usepackage{bm}
\usepackage[inline]{enumitem}
\usepackage{import}
\usepackage[noend]{algpseudocode}
\usepackage[linesnumbered,ruled,vlined, noend]{algorithm2e}

\usepackage[switch]{lineno}

\title{Mixed Q-Functionals: Advancing Value-Based Methods in \\Cooperative MARL with Continuous Action Domains}

\author{Yasin Findik$^*$ and S. Reza Ahmadzadeh% <-this % stops a space
\thanks{PeARL Lab, Richard Miner School of Computer and Information Sciences, University of Massachusetts Lowell, MA, USA {\tt\small yasin\_findik@student.uml.edu, reza@cs.uml.edu}}%
}

\begin{document}

\maketitle

\begin{abstract}

Tackling multi-agent learning problems efficiently is a challenging task in continuous action domains. While \textit{value-based} algorithms excel in sample efficiency when applied to discrete action domains, they are usually inefficient when dealing with continuous actions. \textit{Policy-based} algorithms, on the other hand, attempt to address this challenge by leveraging critic networks for guiding the learning process and stabilizing the gradient estimation. 
The limitations in the estimation of true return and falling into local optima in these methods result in inefficient and often sub-optimal policies. In this paper, we diverge from the trend of further enhancing critic networks, and focus on improving the effectiveness of \textit{value-based} methods in multi-agent continuous domains by concurrently evaluating numerous actions. We propose a novel multi-agent \textit{value-based} algorithm, Mixed Q-Functionals (MQF), inspired from the idea of Q-functionals, that enables agents to transform their states into basis functions. Our algorithm fosters collaboration among agents by mixing their action-values. We evaluate the efficacy of our algorithm in six cooperative multi-agent scenarios. Our empirical findings reveal that MQF outperforms four variants of Deep Deterministic Policy Gradient through rapid action evaluation and increased sample efficiency.

\end{abstract}

\section{Introduction}

In multi-agent reinforcement learning (MARL), the primary goal is to formulate and discover  policies that empower agents to maximize collective or individual rewards in a shared environment. % effectively engage within a shared environment, optimizing collective or individual rewards.
MARL algorithms either incorporate  \textit{policy} gradient~\cite{sutton1999policy} methods for discovering an optimal policy directly or learn \textit{action-value} functions~\cite{watkins1992q} 
for estimating the expected rewards per action in each state, as in single-agent reinforcement learning. Yet the presence of multi-dimensional action spaces reflecting concurrent decisions of all agents poses several challenges such as:
the curse of dimensionality~\cite{shoham2007if}, non-stationarity~\cite{busoniu2008comprehensive}, and the need for a global exploration process~\cite{matignon2012independent}. These challenges are further amplified in scenarios with continuous action spaces, where calculating \textit{each} action's value becomes infeasible.% due to the nature of continuous actions.

Current MARL approaches for continuous action domains primarily utilize \textit{policy-based} methods, such as Multi-Agent Proximal Policy Optimization (MAPPO)~\cite{yu2021surprising}, and Multi-Agent Deep Deterministic Policy Gradient (MADDPG)~\cite{lowe2017multi}. These \textit{policy-based} methods -- mirroring their single-agent counterparts like Proximal Policy Optimization (PPO)~\cite{schulman2017proximal}, Deep Deterministic Policy Gradient (DDPG)~\cite{lillicrap2015continuous} -- are designed to optimize the policy parameters for multiple interacting agents based on expected returns, stabilizing the learning process. Nevertheless, since the \textit{policy-based} methods directly optimize  policies through function approximation, the inherent assumptions and approximations in these methods often lead to sample inefficiency, particularly when contrasted with \textit{value-based} methods~\cite{papoudakis2020benchmarking}.

Meanwhile, recent advances in the development of innovative neural network structures for single-agent RL domains have enabled the \textit{value-based} concept to be applicable in continuous action spaces. The resulting \textit{single-agent} algorithm, known as Q-functionals~\cite{lobel2023q}, has shown to surpass the performance of leading \textit{policy-based} methods. %Meanwhile, recent developments in single-agent learning involve innovative neural network structures that enable \textit{value-based} concept in continuous action spaces, surpassing the performance of leading policy methods, known as Q-functionals~\cite{lobel2023q}.% 
Essentially, Q-functionals convert a state into a function that operates over the action space, facilitating the simultaneous evaluation of Q-values for various actions. This parallel processing capability allows efficient assessment of multiple actions per state, leading to a more effective sampling technique in continuous environments with a \textit{single-agent}.

In this paper, we introduce Mixed Q-Functionals, a novel \textit{value-based} approach tailored to address the sample inefficiency in \textit{policy-based} methods for cooperative multi-agent tasks in continuous action spaces. Unlike frameworks like MADDPG, which rely on a policy network, Mixed Q-Functionals (MQF) enable agents to employ Q-functionals, compute individual action-values efficiently, and mix those to enhance cooperation among multiple agents. We evaluated our approach using six experiments across two different environments, each characterized by cooperative multi-agent tasks.
In summary, the key contributions of this study are:
\begin{enumerate*}[label=(\alph*)]
    \item We introduce Mixed Q-Functionals (MQF) and two baseline methods (Independent Q-Functionals and Centralized Q-Functionals), as novel \textit{value-based} approaches for multi-agent tasks with continuous action domains.

    \item Our research is the first, to the best of our knowledge, to demonstrate the advantages of \textit{value-based} over \textit{policy-based} methods in cooperative MARL with continuous action spaces, revealing new possibilities for \textit{value-based} strategies.

    \item Our comparative analyses reveal that MQF consistently outperforms DDPG-based methods in similar settings, achieving optimal solutions and exhibiting faster convergence in six different scenarios.

\end{enumerate*}

\section{Background and Related Work}

Reinforcement learning (RL) focuses on how an agent can maximize its cumulative future reward, by learning from its interactions within an environment. This concept is commonly modeled, in MARL, as a multi-agent extension of Markov Decision Process (MDP), called Decentralized Partially Observable MDP (DEC-POMDP)~\cite{oliehoek2016concise}. DEC-POMDP is  characterized as a tuple $\langle  \mathcal{S}, \mathcal{A}, \mathcal{O}, \mathcal{R}, \mathcal{T}, N, \gamma\rangle$ where $s \in \mathcal{S}$ indicates the true state of the environment, $\mathcal{A}$, $\mathcal{O}$ and $\mathcal{R}$ are the set of agents' action, observation and reward, respectively. $\mathcal{T} (s, \mathcal{A}, s') \colon \mathcal{S} \times \mathcal{A}\times \mathcal{S} \mapsto [1,0]$ is the dynamics function, defining the transition probability, where $s'$ is the next state. $N$ denotes the number of agents and $\gamma\in[0,1)$ is the discount factor.

\paragraph{Q-Learning \& Deep Q-Networks.} Q-Learning, introduced in ~\cite{watkins1992q} , and Deep Q-Networks (DQN), proposed by ~\cite{mnih2015human}, are two fundamental approaches in RL. Q-Learning utilizes an action-value $Q^{\pi}$ for a policy $\pi$, defined as
\begin{align*}
Q^{\pi}(s, a)=\mathbb{E}[G| s^t=s, a^t=a],
\end{align*}
\noindent where G denotes the return and $t$ is the time-step. This function can be expressed recursively as
\begin{align*}
Q^{\pi}(s, a)=\mathbb{E}_{s'}[R(s,a) + \gamma \mathbb{E}_{a'\sim\pi}[Q^{\pi}(s', a')]],
\end{align*}
\noindent 
with $R$ denoting the reward function. DQN aims to learn the action-value function \smash{$\hat{Q}$} associated with the optimal policy by minimizing the loss:
\begin{align}
\label{loss_DQN}
\resizebox{.91\linewidth}{!}{$
            \displaystyle
            L(w_{\textrm{PN}})=\mathbb{E}_{s,a,r,s'}[(r + \gamma \max_{a'}(\hat{Q}(s', a'; w_{\textrm{TN}}))- \hat{Q}(s, a; w_{\textrm{PN}}))^2],
        $}
\end{align}
\noindent 
where $w_{\textrm{PN}}$ and $w_{\textrm{TN}}$ are the parameters of the prediction and target networks, respectively. The target network parameters $w_{\textrm{TN}}$ are periodically updated with $w_{\textrm{PN}}$ to stabilize learning. To improve stability of the models further, DQN employs an experience replay buffer, storing tuples of $\langle s, a, r, s'\rangle$.

The direct application of Q-Learning to multi-agent scenarios might involve each agent $i$ independently learning an optimal function \smash{$\hat{Q}_i^\ast$}, known as Independent Q-Learning (IQL)~\cite{tan1993multi}. Yet, as each agent adjusts its policy autonomously over time, the environment becomes non-stationary from the perspective of other agents, disrupting the stationarity conditions necessary for the Q-Learning algorithm to reliably converge. Another approach is fully centralized learning, where all agents share a single controller and learn a value function together~\cite{claus1998dynamics}. However, this method is computationally expensive and can become intractable since the observation and action space grow exponentially with the number of agents.
\paragraph{Value Function Factorization.}Value function factorization methods are proposed as an efficient approach for learning a joint action-value function, whose complexity increases exponentially with the number of agents. These methods adhere the centralized training with decentralized learning (CTDE)~\cite{oliehoek2008optimal} paradigm, enabling agents to execute their actions independently while a centralized mechanism integrates their strategies. Thus, they effectively address the non-stationarity problem of IQL through centralized training, and overcome the scalability issue of centralized learning through decentralized execution. 

VDN~\cite{sunehag2017value} and QMIX~\cite{rashid2020monotonic} are two exemplary methods for factorizing value functions. Both methods maintain a separate action-value function \smash{$\hat{Q}_i$} for each agent $i \in \{ 1,...,N\}$. Their difference lies in 
how they compute the central action value
$Q_{\textrm{tot}}$. Specifically, VDN sums $Q_i$s to obtain $Q_{\textrm{tot}}$, as 
\begin{align}
\label{VDN}
\hat{Q}_{\textrm{tot}} \coloneqq \sum_{i=1}^{N} \hat{Q}_i(s, a_i),
\end{align}
while QMIX mixes them using a continuous monotonic function which is state-dependent, as follows:
\begin{align}
\label{QMIX}
\hat{Q}_{\textrm{tot}} \coloneqq f_s(\hat{Q}_1(s, a_1), ..., \hat{Q}_n(s, a_N)),  
\end{align}
where $\frac{\partial f_s}{\partial \hat{Q}_i} \ge
0,  \forall i \in \{1, ..., N\}$. These value function factorization methods most commonly utilize DQN to approximate the action value function. They minimizes loss \eqref{loss_DQN} in a centralized manner, by substituting \smash{$\hat{Q}$} with \smash{$\hat{Q}_{\textrm{tot}}$}, and $r$ with $r_{\textrm{team}}$ which is the sum of agents' rewards.

\textit{Value-based} methods are effective in deriving robust policies for agents in environments with a finite number of actions (i.e. discrete actions). Yet, these methods, including their enhanced versions~\cite{findik2023impact}, often struggle to generate viable policies in continuous environments, where the action space is represented by continuous vectors, resulting in an infinite array of possible actions~\cite{lim2018actor}.

\paragraph{Policy Gradient (PG).}
Policy Gradient (PG) methods, as described in~\cite{sutton1999policy}, are pivotal in RL with continuous action spaces. They directly optimize the policy parameters (denoted as  $\theta$), aiming to maximize the expected return. The core principle involves adjusting  $\theta$ in the direction of the policy's gradient, expressed as:
\begin{align}
\label{PG}
\nabla_\theta J (\theta)= \mathbb{E}_{s \sim p^{\pi}, a \sim \pi_{\theta}}[\nabla_\theta\log \pi_{\theta} (a|s)Q^\pi(s,a)],
\end{align}
\noindent
where $p^{\pi}$ represents the state distribution under policy $\pi$. Diverse algorithms have emerged from PG, each characterized by unique approaches to estimating $Q^\pi$. The REINFORCE algorithm, for instance, employs a sample return calculation $G_t = \sum_{k=1}^{\infty} \gamma^{k-1}G_{t+k}$ as its foundational mechanism~\cite{williams1992simple}. Meanwhile, temporal-difference methods are employed in actor-critic algorithms to learn an action-value function approximation \smash{$\hat{Q}^\pi(s, a)$}, which serves as the critic \cite{konda1999actor}.

The PG theorem can also be adapted for deterministic policies (DPG) ~\cite{silver2014deterministic}, denoted as $\mu_{\theta}: \mathcal{S} \mapsto \mathcal{A}$ with a policy vector of $n$ parameters $\theta \in \mathbb{R}^n$. The gradient of the expected return can be formulated as: 
\begin{align}
\label{DPG}
\nabla_\theta J (\mu_\theta)= \mathbb{E}_{s \sim p^{\mu}}[\nabla_\theta\mu_{\theta}(s) \nabla_{a}Q^{\mu}(s,a)|_{a=\mu_{\theta}(s)}].   
\end{align}
\noindent
Deep Deterministic Policy Gradient (DDPG)~\cite{lillicrap2015continuous}, a variation of DPG, uses deep neural networks to approximate the policy $\mu$ and the critic \smash{$\hat{Q}^\mu$}. 

\paragraph{Multi-Agent DDPG (MADDPG).}MADDPG~\cite{lowe2017multi} is an actor-critic \textit{policy-based} method built upon DDPG, and designed to operate with multiple agents. In this approach, agent policies parameterized by $\bm{\theta} = [\theta_1, \dots, \theta_n]$ and denoted as $\bm{\pi}=[\pi_1, \dots, \pi_N]$. The gradient of the expected return, shown in~\eqref{PG}, for each agent is re-expressed as:
\begin{align*}
\resizebox{.95\linewidth}{!}{$
            \displaystyle
            \nabla_{\theta_i} J (\theta_i)= \mathbb{E}_{x \sim p^{\mu}, a_i \sim \pi_{i}}[\nabla_{\theta_i} \log \pi_{i} (a_i|o_i)\hat{Q}_i^{\bm{\pi}}(\bm{o}, \bm{a})],
        $}
\end{align*}
\noindent
where \smash{$\hat{Q}_i^\pi(\bm{o}, \bm{a})$} is a \textit{centralized} action-value function that takes state information $\bm{o}=[o_1,\dots,o_N]$ and the actions of all agents $\bm{a}=[a_1,\dots,a_N]$ as input and outputs the $Q$-value. Next, the idea is extended to work with deterministic policies $\bm{\mu}=[\mu_{\theta_1}, \dots, \mu_{\theta_N}]$, and \eqref{DPG} is updated as follows:
\begin{align*}
\resizebox{.95\linewidth}{!}{$
            \displaystyle
            \nabla_{\theta_i} J (\mu_{\theta_i}) = \mathbb{E}_{\bm{o} \sim p^{\bm{\mu}}, \bm{a}\sim \bm{\mu}}[\nabla_{\theta_i}\mu_{\theta_i}(a_i|o_i) \nabla_{a_i}\hat{Q}_i^{\bm{\mu}}(\bm{o}, \bm{a})|_{a_i=\mu_{\theta_i}(o_i)}].
        $}
\end{align*}
Like DQN, MADDPG also uses replay memory, which keeps the experiences of all agents, and target networks to increase the stability of the policy networks. Including these features, the loss, utilized to update the centralized action-value function $Q_i^{\bm{\mu}}$, becomes as follows:
\begin{align*}
\resizebox{.98\linewidth}{!}{$
            \displaystyle
            L(\theta_{i})=\mathbb{E}_{\bm{o},\bm{a},\bm{r},\bm{o'} \sim \mathcal{D}}[(r_i + \gamma Q_i^{\bm\mu'}(\bm{o'},\bm{a'} )|_{\bm{a'}=\bm\mu'(\bm{o'})} - Q_i^{\bm\mu}(\bm{o},\bm{a} ))^2], 
        $}
\end{align*}
\noindent
where $\mathcal{D}$ is the replay memory that contains the tuples $\langle\bm{o}, \bm{a}, \bm{r}, \bm{o'}\rangle$, and $\bm{\mu'} = [\mu_{\theta'_1}, \dots, \mu_{\theta'_N}]$ is target policies whose parameters are regularly updated with $\bm{\theta}$ (for detail explanation, see~\cite{lowe2017multi}). Although MADDPG-based algorithms perform state-of-the-art by maximizing agents' return for higher rewards, in complex environments (i.e. environments with large state, and motion space), the results of training tend to fall into local optimal solutions due to inefficient sampling and their small scope of exploration.

Recent studies have devoted efforts to enhancing the sample efficiency and overall performance of actor-critic based PG methods. Notably, advancements such as FACtored Multi-Agent Centralized policy gradients (FACMAC)~\cite{peng2021facmac} and Value Function Search (VFS)~\cite{marchesini2023improving} have emerged as significant contributions. FACMAC, for instance, incorporates a factored critic and a centralized gradient estimator to facilitate learning in continuous cooperative tasks. VFS introduces a novel approach by periodically generating a suite of perturbed critics through a two-scale perturbation noise, specifically designed to refine value predictions for a given Deep PG agent, which utilizes actor-critic networks. However, our research diverges by focusing on applying the \textit{value-based} concept in continuous MARL tasks, addressing the inherent inefficiencies in sampling commonly associated with \textit{policy-based} methods.

\section{Proposed Method}

In the preceding section, we discussed the challenges and advantages of policy-based and value-based methods in multi-agent settings. Policy-based methods are effective in continuous action spaces, but often converge to suboptimal solutions or show slow convergence due to sample inefficiency. To address these problems, we propose a novel value-based MARL algorithm, called Mixed Q-Functionals (MQF). MQF leverages sample efficiency of Q-functionals~\cite{lobel2023q} for continuous environments, and employs CTDE paradigm to tackle issues of scalability and non-stationarity. Additionally, we introduce Independent Q-Functionals (IQF) and Centralized Q-Functionals (CQF) to demonstrate the superior performance of MQF in multi-agent contexts.

Since our proposed MARL algorithms build upon the foundation of QF, we offer a brief overview of it before delving into the specifics of IQF, CQF, and MQF.

\subsection{Q-functionals}
Q-functionals~\cite{lobel2023q} rely on the main idea that if the values of numerous actions can be calculated quickly, then value-based methods, known for their sample efficiency, become suitable for continuous action spaces. This fundamental idea allows to address the shortcomings of policy-based methods effectively for \textit{single-agent} setups. The concept of Q-functionals modifies the representation of Q-functions to enable the simultaneous evaluation of multiple action-values for a specific state. Traditional deep Q-functions, denoted by a neural network, map state-action pairs directly to $\mathcal{R}$, as in
\begin{align}
\label{tqa}
\hat{Q}(s,a): (\mathcal{S} \times \mathcal{A}) \mapsto \mathcal{R}.
\end{align}
\noindent
However, Q-functionals redefine this mapping by separating the state and action, where the state alone is mapped to a function that subsequently maps actions to $\mathcal{R}$:
\begin{align}
\label{fqa}
\hat{Q}^{\textrm{F}}(s,a): \mathcal{S} \mapsto (\mathcal{A} \mapsto \mathcal{R}).
\end{align}

\begin{figure*}[t]
\centering
  \includegraphics[width=\linewidth]{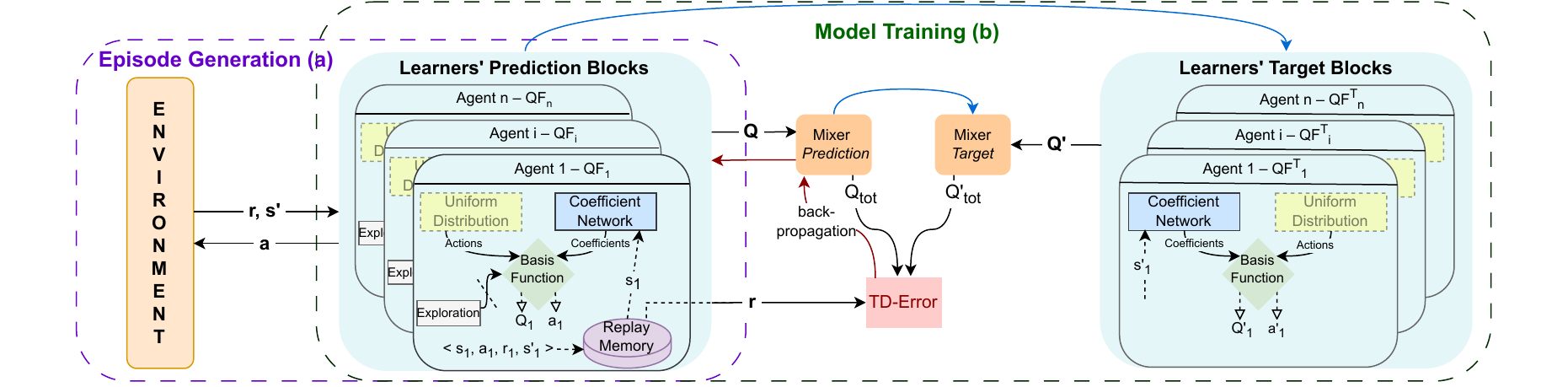}
  \caption{\small{Overview of the MQF Architecture: (a) Episode Generation, (b) Model Training. The blue arrows depict the target network updates for mixer and learners' prediction blocks, while the red arrows indicate the direction of the backpropagation.
  }} 
    \label{fig:system_architecture}
\end{figure*}

The standard architectures, as described by~\eqref{tqa}, require twice the computational effort to evaluate action-values for two actions in the same state compared to just one. Yet, Q-functionals transform each state into a set of parameters defining a function over the action space. This means each state is encoded as a function, constructed by learning the coefficients of basis functions within the action space. These functions, representing states, enable quick evaluation of various actions through matrix operations between the action representations and learned coefficients. Hence, Q-functionals efficiently handle environments with continuous action spaces for \textit{single-agent} settings without sacrificing sample efficiency.

\subsection{Independent Q-Functionals (IQF)}
\label{body:IQF}

We introduce a straightforward application of Q-functionals in multi-agent context, wherein each agent $i$ independently aims to learn an optimal functional, \smash{${\hat{Q}}^{\textrm{F}*}_i$} in~\eqref{fqa}, similar to the concept underlying IQL~\cite{tan1993multi}. However, this approach also cannot handle the non-stationarity that emerges due to the evolving policies of the learning agents. Therefore, it lacks convergence guarantees, even with unlimited exploration. In our study, we use Independent Q-Functionals (IQF) as a baseline, evaluating the efficacy of Q-functionals variation against DDPG-based methods.

\subsection{Centralized Q-Functionals (CQF)}
\label{body:CQF}

In our study, we also present another variant of the Q-functionals for multi-agent settings through the introduction of a single Centralized Q-Functionals (CQF), denoted as \smash{${\hat{Q}}^{\textrm{F}}_{\textrm{cent}}$}, which integrates the state functions through learned coefficients over the joint action space of all agents. This approach can be expressed by modifying \eqref{fqa} as follows:
\begin{align}
\hat{Q}^{\textrm{F}}_{\textrm{cent}}(\bm{s},\bm{a}): \mathcal{S} \mapsto (\mathcal{A} \mapsto \mathcal{R}),
\end{align}
\noindent
where $\bm{a}=[a_1,\dots,a_N]$ and $\bm{s}=[s_1,\dots,s_N]$ represent actions and states of all agents, respectively. Nevertheless, as the number of agents grows, the size of the state and action spaces increases exponentially, leading to a rise in the model's time complexity. Consequently, maintaining sampling efficiency becomes more challenging.

\subsection{Mixed Q-Functionals (MQF)}
\label{body:MQF}

In response to the non-stationarity challenges inherent in IQF and the scalability issues faced by CQF, we introduce Mixed Q-Functionals (MQF). This novel method mixes Q-values derived from each agent's individual Q-functionals. It utilizes a centralized loss function to train the agents' coefficient networks, leveraging the concept of value function factorization. The mixing function in MQF is versatile, ranging from additive forms as in \eqref{VDN} to monotonic functions like in \eqref{QMIX}, or more complex methods, similar to those described in~\cite{son2019qtran}~\cite{rashid2020weighted}. The complexity of the tasks being addressed plays a crucial role in determining the specific form of the mixing function to be employed.

Notably, MQF differentiates itself from other \textit{value-based} MARL methods, described in current literature on mixing functions, through its adept handling of continuous action spaces. It also capitalizes on the sample efficiency characteristic of \textit{value-based} concept, enhancing agent performance relative to \textit{policy-based} methods. MQF\footnote[1]{The pseudo-code is detailed in Supplementary Material~\ref{MQF}.
%~\ref{MQF}.
} framework enables agents to represent their states with basis functions, facilitating its application in environments with continuous action spaces. This functionality, combined with the mixing of computed action-values, promotes cooperative behavior among agents in such settings. The effectiveness of MQF is underscored in the results section, where it is benchmarked against DDPG variants, specifically Independent DDPG (IDDPG), Centralized DDPG (CDDPG), and MADDPG. %The pseudo-code of MQF can be found in Supplementary Material~\ref{alg:MQF}

As illustrated in Fig.~\ref{fig:system_architecture}, 
our proposed framework entails three main components: generating episodes, training learners' prediction blocks, and updating learners' target blocks.

\paragraph{Generating Episodes.} 
The generation of an episode involves agents engaging with the environment through actions derived from their prediction blocks, as illustrated in Fig.~\ref{fig:system_architecture}(a). Concisely, the process starts with the agent's coefficient neural network predicting the basis function coefficients for the current state. This is followed by uniformly sampling and evaluating $k$ actions, which involves calculating the matrix multiplication of the predicted coefficients and the sampled actions' representations. The action associated with the highest $Q$-value is, then, selected. To enhance the exploration and sample efficiency, in our experiments, we employ \textit{Gaussian} exploration, by adding \textit{Gaussian}-distributed noise to agents' calculated actions, or $\varepsilon$-greedy, where agents may take random actions with $\varepsilon$ probability. Notably, these exploration techniques are deactivated during agent evaluation. Upon receiving a vector of actions, $\bm{a}$, from the agents' prediction blocks, the environment provides feedback in the form of $\bm{r}$ and $\bm{s}$, representing the agents' rewards and states, respectively. These vectors are subsequently allocated to the corresponding agents, and stored in their replay memories, $\mathcal{D}_i$, as tuples $\langle s_i, a_i, r_i, s_i' \rangle$.

\paragraph{Training Learners' Prediction Blocks.}
The training process for the MQF begins with the selection of a batch of transition tuples $\langle s_i, a_i, r_i, s_i'\rangle$, each of size $b$, from the replay memory $\mathcal{D}_i$ of each agent. These tuples are utilized to calculate the agents' individual $Q_i$ values. Subsequently, we employ two networks: one for mixing the agents' current $Q$ values to derive $Q^{\textrm{F}}_{\textrm{tot}}$, and another for mixing the next-state $Q'$ values to obtain $Q'^{\textrm{F}}_{\textrm{tot}}$. Fig.~\ref{fig:system_architecture} illustrates these networks as the mixer prediction and the mixer target, respectively. This procedure is followed by the calculation of the temporal difference error between the prediction and target values, which can be encapsulated by the following formulation:
\begin{align}
\label{e_td}
e_{\textrm{TD}} &= \sum_{\bm{s}, \bm{a}, \bm{r}, \bm{s}' }^{\bm{b} \sim \bm{\mathcal{D}}} [(\hat{Q}^{\textrm{F}}_{\textrm{tot}}(\bm{s}, \bm{a}; \bm{w}_{\textrm{PN}}) - y(\bm{r},\bm{s}'))^2],
\end{align}
\noindent where $y(\bm{r},\bm{s}')$ is defined as:
\begin{align}
\label{y}
y(\bm{r},\bm{s}') &=\sum_{i=1}^{N} \bm{r}_i + \gamma \max_{\bm{a}'}(\hat{Q}^{\textrm{F}}_{\textrm{tot}}(\bm{s}', \bm{a}'; \bm{w}_{\textrm{TN}})|_{\bm{a}'\sim \mathcal{U}}).
\end{align}
\noindent  
The bold symbols, in \eqref{e_td} and \eqref{y}, are vectors denoting the set of corresponding values for all agents and $\mathcal{U}$ signifies a uniform distribution. The function \smash{$\hat{Q}^{\textrm{F}}_{\textrm{tot}}$}, resembling forms such as those in~\eqref{VDN} and~\eqref{QMIX}, serves as a mixing mechanism for the agents' $Q$-values. The optimization of the agents' Q-functionals aims to minimize the temporal difference error, $e_{\textrm{TD}}$ in \eqref{e_td}, thus facilitating the coordination of agent actions to maximize collective rewards. This training is conducted centrally using \smash{$\hat{Q}^{\textrm{F}}_{\textrm{tot}}$}, while the determination of individual agent actions is guided by their \smash{$\hat{Q}^{\textrm{F}}_i$}, enabling decentralized execution in line with the CTDE paradigm. The comprehensive schematic of the training process is depicted in Fig.~\ref{fig:system_architecture}(b).

\paragraph{Updating Learners' Target Blocks.}

Our framework, designed for multi-agent environments with continuous action spaces, diverges from standard value-based methods in updating the target networks. It employs a \textit{soft} update mechanism at each time-step, proven more effective in continuous action domains instead of \textit{periodic} updates~\cite{lillicrap2015continuous}. This method is characterized by incremental adjustments, formalized as $w_{\textrm{TN}} = \tau w_{\textrm{PN}} + (1-\tau)w_{\textrm{TN}}$ where $\tau$ is a small factor ($\tau \ll 1$), ensuring improved stability and effectiveness in such environments.

\section{Experiments}

\subsection{Environments}
\label{body:environments}

We evaluate our method in two distinct MARL environments, Multi-Agent Particle Environment~\cite{mordatch2017emergence} and Multi-Walker Environment~\cite{terry2021pettingzoo}, across various scenarios\footnote[2]{Supplementary Material~\ref{environments}
%~\ref{environments} 
includes images of the scenarios.}.

\subsubsection{Multi-Agent Particle Environment (MPE)}
\label{section:MPE}
Utilizing OpenAI's MPE package~\cite{lowe2017multi}, we devise two novel scenarios: landmark capturing and predator-prey, each with two variants. The action space, in these scenarios, is two-dimensional governing both vertical and horizontal movements, within a range of $[-1, 1]$.

\paragraph{Landmark Capturing:}
This scenario involves $N$ agents and $N$ landmarks, with agents cooperating to cover all landmarks. An agent's reward is \smash{$e^{-d^2/c}$}, where $d$ is the \textit{Euclidean} distance to the nearest landmark within a threshold $d_{th}$, or the sum of distances to all landmarks otherwise, leading agents to strategize which landmark is to be captured by whom. Each agent's observation includes its velocity and landmarks' relative  location. We explore two configurations: $2$ agents with $2$ landmarks (2A2L) and $5$ agents with $5$ landmarks (5A5L).

\paragraph{Predator-Prey (PP):}
In this scenario, we adapt the \textit{simple tag} environment, initially introduced by~\cite{lowe2017multi}, to a variation of the classic predator-prey game. This setup features three slower agents (predators) aiming to catch a faster prey in a 2D space with two large landmarks serving as obstacles. The prey's movements, based on a hard-coded heuristic, aim to maximize distance from the nearest predator at each step. Agents have access to sensory information: their own velocity and position, relative position of landmarks, other agents, the prey, and the prey's velocity. Two variants, differentiated by their reward structures, are implemented:

\begin{itemize}
    \item \textit{Standard Predator-Prey (S-PP)}: 
    Success is achieved when at least two predators collide with the prey, with each predator involved gaining $10$ points. A distance-based penalty (proportional to the \textit{Euclidean} distance between predator and prey) is applied otherwise. This variant, however, leads predators to unintentionally cooperate to collide with the prey due to distance-based penalization.

    \item \textit{Increased Cooperation Predator-Prey (IC-PP)}: 
    This variant introduces a field of view mechanic for predators, promoting more strategic and intentional cooperation. A successful capture necessitates at least one predator physically colliding with the prey. Predators that have the prey within their field of view at the moment of capture gain $10$ points, encouraging them to time the capture to maximize team reward.
    %Predators are rewarded with $10$ points when within the prey's field of view at the moment of capture, encouraging them to time the capture to maximize team reward.
    
\end{itemize}

\begin{table*}[bp]
\centering
\resizebox{\textwidth}{!}{
\begin{tabular}{ccc|cc|ccc|ccc}
\hline
            Methods & \multicolumn{2}{c|}{2-Agent \& 2-Landmarks} & \multicolumn{2}{c|}{5-Agents \& 5-Landmarks} & \multicolumn{3}{c|}{Standard Predator-Prey} & \multicolumn{3}{c}{Increased Cooperation Predator-Prey} \\
            & Team Reward    & Success Rate     & Team Reward     & Success Rate     & Team Reward     & Total \#captures    & 3-Agent \#captures    & Team Reward     & Total \#captures    & 3-Agent \#captures   \\
CQF         & 32.64 ± 4.26         & 0.09 ± 0.01                       & 6.22 ± 0.54         & 0.00 ± 0.00                       & 27.44 ± 0.66         & 1.82 ± 0.03               & 0.00 ± 0.00            & 65.27 ± 1.60         & 4.80 ± 0.11               & 0.37 ± 0.03           \\
CDDPG       & 25.86 ± 4.98        &  0.00 ± 0.00                           & 5.89 ± 0.32         & 0.00 ± 0.00                         & 20.13 ± 1.08         & 1.51 ± 0.05               & 0.00 ± 0.00            & 72.77 ± 5.35         & 4.30 ± 0.11               & 1.04 ± 0.21           \\
IQF         & 74.98 ± 2.13         & 0.69 ± 0.07                         & 207.67 ± 0.49         & 0.94 ± 0.02              & \textbf{47.85 ± 0.47}              & \textbf{2.73 ± 0.02}         & \textbf{0.10 ± 0.00}               & 97.61 ± 1.13            & \textbf{4.91 ± 0.06}         & 2.03 ± 0.12                          \\
IDDPG       & 77.38 ± 0.29        & 0.83 ± 0.01                        & 169.55 ± 5.93        & 0.13 ± 0.11                         & 36.69 ± 0.48         & 2.22 ± 0.03               & 0.07 ± 0.00            & 86.43 ± 0.30         & 4.19 ± 0.02               & 1.91 ± 0.01           \\
MQF         & \textbf{80.46 ± 0.57}        & \textbf{0.88 ± 0.03}                        & \textbf{210.69 ± 0.32}         & \textbf{0.98 ± 0.01}                        & 33.97 ± 0.58        & 2.02 ± 0.03               & 0.07 ± 0.00            & \textbf{104.02 ± 0.96}        & 4.65 ± 0.05               & \textbf{2.71 ± 0.04}           \\
MADDPG-Ind  & 75.05 ± 3.08         & 0.82 ± 0.08                         & 176.28 ± 5.09        & 0.29 ± 0.16                       & 34.11 ± 0.11         & 2.05 ± 0.00               & 0.06 ± 0.00            & 81.92 ± 0.86         & 3.78 ± 0.05               & 2.11 ± 0.03           \\
MADDPG-Team & 73.90 ± 3.17        & 0.82 ± 0.08                       & 166.25 ± 6.64         & 0.02 ± 0.01                         & 38.13 ± 0.28         & 2.25 ± 0.01              & 0.07 ± 0.00            & 78.87 ± 0.79         & 3.75 ± 0.70               & 1.93 ± 0.01           \\ \hline

\end{tabular}
}

\caption{
Average metrics with 95\% confidence intervals for multiple runs on MPE with different scenarios upon training completion.
}
\label{table:experiment_results_mpe}

\end{table*}

\subsubsection{Multi-Walker Environment (MWE)} 

The MWE, a multi-agent continuous control task emphasizing locomotion, was initially introduced in~\cite{gupta2017cooperative}. It consists of walkers, each controlling two joints in a leg, tasked with transporting objects atop them. Originally, each walker is considered a single agent, focusing on coordinating movements to balance and transport a package. Our study, however, introduces a novel modification: splitting a walker into two agents, each controlling one leg, to include additional cooperative behaviors. This change requires each leg of a walker to not only coordinate internally but also to cooperate with agents controlling the legs of others.

Each agent (i.e., a leg) in this modified environment controls two torque values, ranging from $[-1, 1]$. While the original observation space per agent includes a 32-dimensional vector with information about nearby walkers and LiDAR sensor data, our version modifies it by including global (hull's angle, angular velocity, package angle, velocities in x and y directions) and local (ground contact, hip and knee angles, and velocities) information about the walker and leg, respectively. Agents lose $100$ points for falls but gain $5$ points for successful forward movement. The environment terminates an episode under any of the following conditions: a walker or package falls, reaches the edge of the terrain, or a predefined time-step is met. We utilize two configurations: one walker partitioned into two agents (2A1W) and two walkers partitioned into four agents (4A2W).

\subsection{Models and Hyperparameters}

Our algorithm approximates the Q-functionals for each agent using a Multi-Layer Perceptron (MLP) with a variable number of layers. These networks take a state as input and output a number of coefficients determined by the combinatorial formula ${r + d \choose d}$, where $d$ is the action size and $r$ represents the order of basis function used for state representation. We used \textit{polynomial} basis functions of rank $r=2$ for CQF, IQF, and MQF in our experiments. For MADDPG, we utilized its original implementation\footnote[3]{https://github.com/openai/maddpg}.
%For the MADDPG, we utilize the original implementation as provided here\footnote[3]{https://github.com/openai/maddpg}.
We determined optimal hyperparameters\footnote[4]{Hyperparameters are detailed in Supplementary Material~\ref{experimental_details}.
%~\ref{experimental_details}.
} through testing across two different seeds.

Our results are presented through figures -- Fig.~\ref{fig:body_landmark-capturing}, Fig.~\ref{fig:body_predator-prey} and Fig.~\ref{fig:body_multi-walkers} -- that display mean test rewards of agents, indicated by solid lines, with shaded regions representing the $95\%$ confidence interval calculated over multiple runs. Test rewards are obtained by evaluating agents' rewards using a greedy strategy, with assessments conducted every $10000$ training steps over $10$ test episodes. Comprehensive numerical results are summarized in Table~\ref{table:experiment_results_mpe} and Table~\ref{table:experiment_results_walker}, based on the average of $10000$ test episodes after training. The implementation and experimental setup are available on GitHub\footnote[5]{Link to be provided upon acceptance.}.

\subsection{Results}

We evaluate the performance of our method on six distinct tasks, as explained in Section~\ref{body:environments}.
In each of these tasks, we assess the performance of our proposed algorithm, MQF (Section~\ref{body:MQF}), along with its simpler extensions -- CQF and IQF (Sections~\ref{body:CQF} ands~\ref{body:IQF}, respectively) -- against established benchmarks: Centralized DDPG (CDDPG), Independent DDPG (IDDPG), and Multi-Agent DDPG (MADDPG). It should be noted that we employ two variants of the MADDPG: MADDPG-Team, trained with team-based rewards, and MADDPG-Ind, using individual agent rewards. 

\begin{figure}[t]
\centering
  \includegraphics[width=\linewidth]{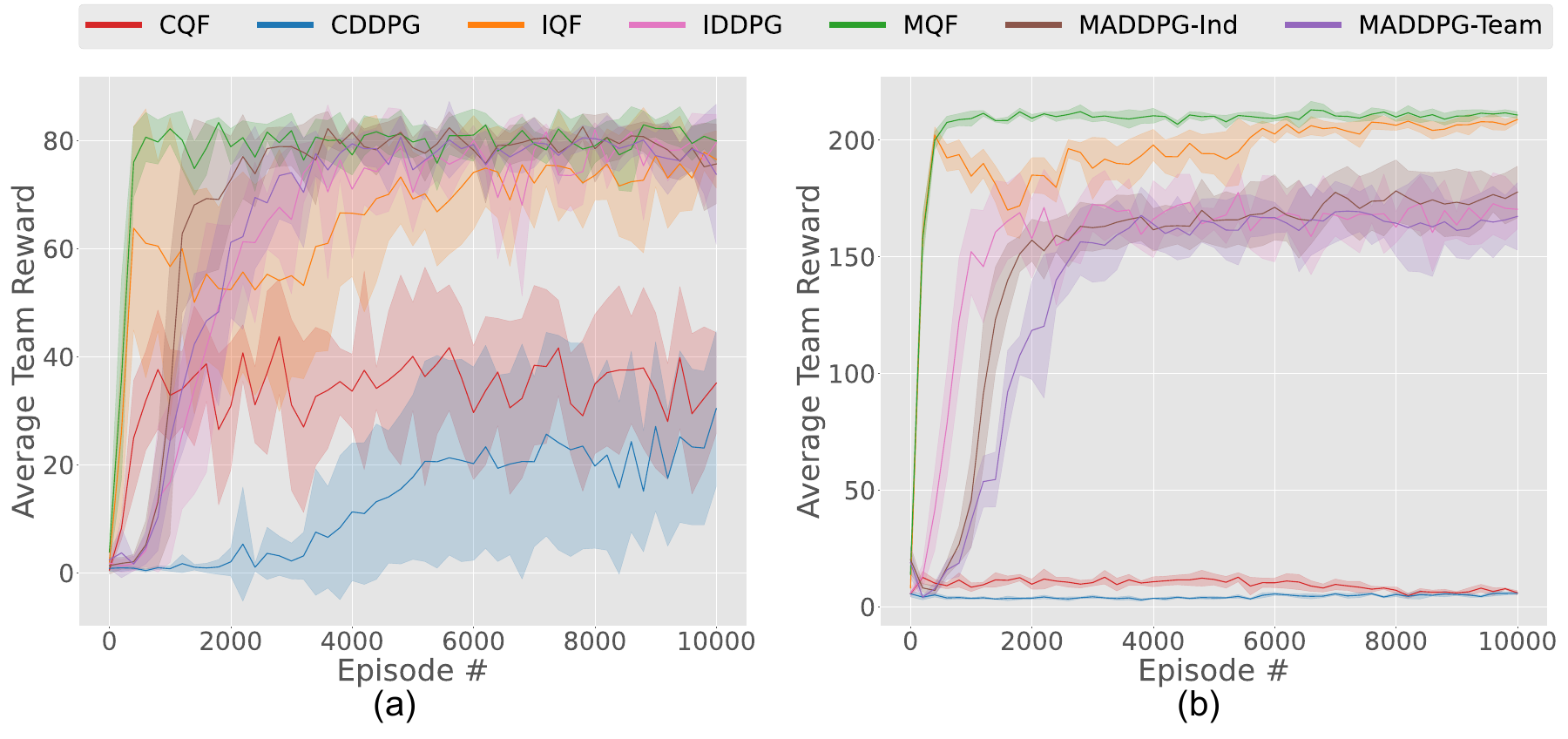}
  \caption{\small{Landmark Capturing Results: (a) 2A2L, (b) 5A5L.}} 
    \label{fig:body_landmark-capturing}
\end{figure}

In our experimental analysis, we begin with the MPE, specifically chosen as the MADDPG algorithm was originally evaluated with it, providing a direct basis for comparison. The first experiments within MPE focus on the landmark capturing scenarios (2A2L and 5A5L).
The observations derived from Fig.~\ref{fig:body_landmark-capturing} reveal that regardless of the agent count, the centralized methods, CQF and CDDPG, exhibited subpar performances. This can be attributed to two primary factors: (i) the curse of dimensionality resulting from the concatenation of agent observations, and (ii) the limited nature of agent observations, as previously discussed in Section~\ref{section:MPE}. Specifically, agents only perceive their velocity and the relative location of landmarks, without awareness of their own positions, leading to an ineffective distribution of landmarks among agents. Notably, while CQF and CDDPG exhibit improved performances in the 2A2L scenario compared to 5A5L, they, nonetheless, underperform relative to other approaches.

On a different note, the two MADDPG variants and independent learning models, IQF and IDDPG, demonstrate slower convergence to \textit{any} solution than MQF in both 2A2L and 5A5L setups, as demonstrated in Fig.~\ref{fig:body_landmark-capturing}(a) and Fig.~\ref{fig:body_landmark-capturing}(b), respectively. A solution is considered optimal if all the landmarks are captured by the agents, and otherwise suboptimal. While all algorithms exhibit comparable performances in the 2A2L (i.e., converging to an \textit{optimal} solution); in the 5A5L setup, a noticeable difference is observed. DDPG-based algorithms tend to converge to suboptimal solutions, likely due to sample inefficiency, whereas value-based methods, including IQF, show a tendency to converge to optimal solutions. Notably, IQF exhibits challenges in addressing the non-stationarity issues of independent learning. This is particularly evident from the performance fluctuations observed around the $2000$th episode mark, although its performance does show improvement with extended training.
In contrast, MQF which utilizes CTDE paradigm to tackle these problems, rapidly converges to an optimal solution. Furthermore, we defined a success metric for these experiments: an episode is successful if all landmarks are captured by its end.
As depicted in Table~\ref{table:experiment_results_mpe}, MQF outperforms in both reward and success rate, proving its effectiveness in these scenarios.

\begin{figure}[t]
\centering
  \includegraphics[width=\linewidth]{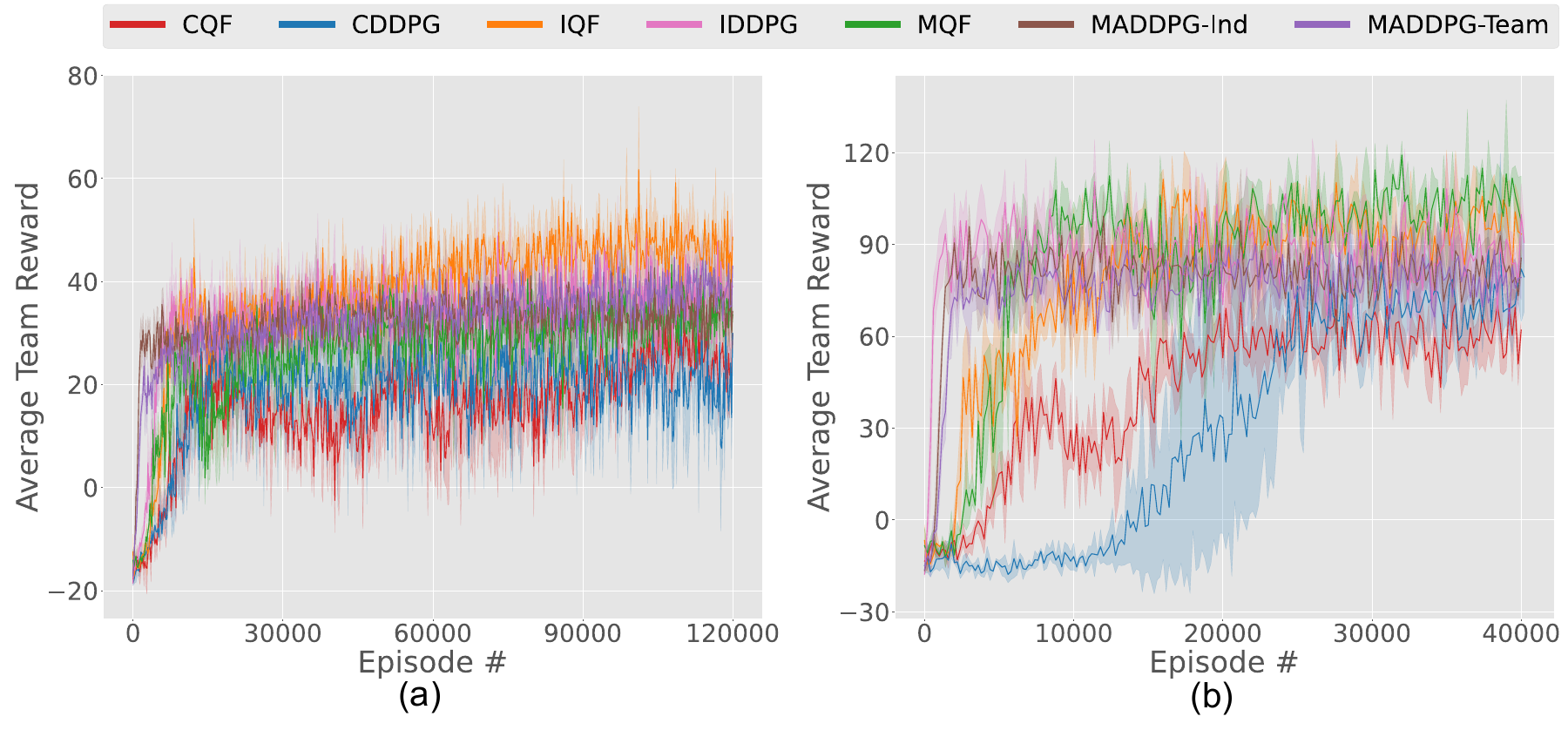}
  \caption{\small{Predator-Prey Results: (a) S-PP, (b) IC-PP.}} 
    \label{fig:body_predator-prey}
\end{figure}

In the second phase of MPE experiments, we concentrate on predator-prey scenarios, specifically S-PP and IC-PP. As demonstrated in Fig.~\ref{fig:body_predator-prey}(a) and detailed in Table~\ref{table:experiment_results_mpe}, the S-PP scenario showcases the superiority of the IQF algorithm over others, including MQF, with an average reward of $47.85 \pm 0.47$. Interestingly, despite the fact that the agents' collective reward per capture increases when three agents concurrently catch the prey, none of the algorithms consistently converge to this behavior. This is because securing multiple captures within an episode yields a higher cumulative reward in the long term. To elucidate, a single capture involving three agents yields a team reward of $30$. However, this strategy is less time-efficient compared to capturing with two agents, which results in a reward of $20$ but allows for more frequent captures. Evidently, as the number of three-agent captures rises, the total number of captures diminishes, leading to a decrease in the overall episode team reward. The additional $10$ reward does not sufficiently incentivize collaboration, suggesting that this scenario is more effectively resolved by agents operating independently rather than increasing complexity through mixing agents action-values. This experiment underscores that independent tasks within a multi-agent setting can be effectively addressed using Q-functionals, as evidenced by its performance relative to DDPG-based approaches.

\begin{figure}[b]
\centering
  \includegraphics[width=\linewidth]{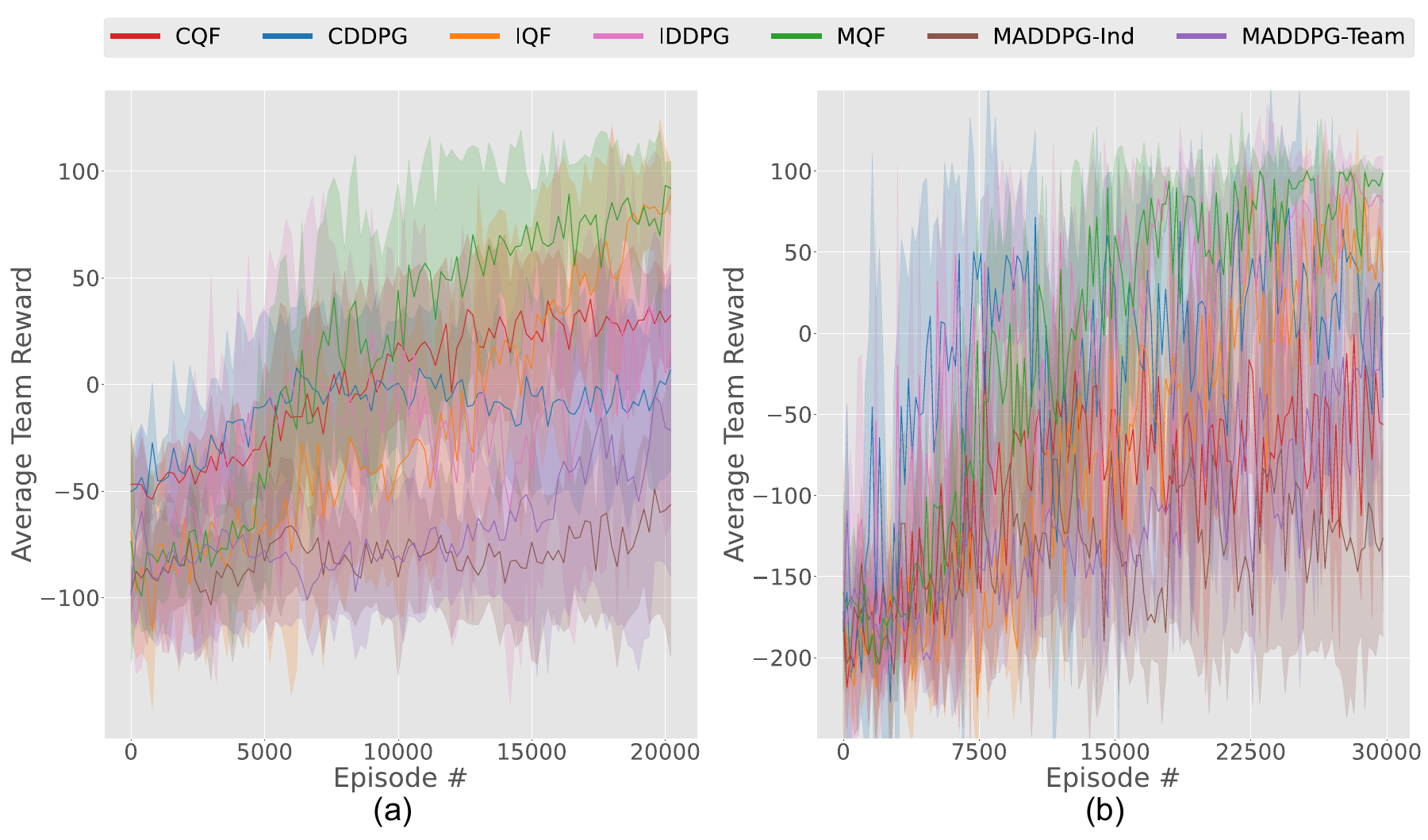}
  \caption{\small{Multi-Walker Results: (a) 2A1W, (b) 4A2W.}} 
    \label{fig:body_multi-walkers}
\end{figure}

However, in the IC-PP scenario, we demonstrate MQF's effectiveness in promoting cooperative behaviors among agents. This scenario rewards agents not only for capturing prey but also for observing it during capture, encouraging cooperative capture strategies. 
As shown in Table~\ref{table:experiment_results_mpe}, although IQF achieves a higher total number of prey captures, MQF outperforms both IQF and DDPG extensions in terms of average team reward.
This is due to MQF's proficiency in facilitating three-agent captures, leading to enhanced cooperative behaviors and higher collective rewards. Figure~\ref{fig:body_predator-prey}(b) indicates MADDPG-Team and MADDPG-Ind converge more rapidly but to local optima, making them less effective than IQF and MQF in maximizing team rewards, as highlighted in Table~\ref{table:experiment_results_mpe}.

\begin{table}[t]
\centering
\resizebox{\linewidth}{!}{
\begin{tblr}{
  cells = {c},
  cell{1}{2} = {c=2}{},
  cell{1}{4} = {c=2}{},
  vline{3} = {1}{},
  vline{4} = {1-9}{},
  hline{1,10} = {-}{},
}
            & 2-Agents (1-Walker) &              & 4-Agents (2-Walkers) &              \\
            & Team Reward    & Balance Rate & Team Reward    & Balance Rate \\
CQF         & 50.76 ± 13.09        & 0.72 ± 0.07            & -59.51 ± 8.15        & 0.46 ± 0.03\\
CDDPG       & -0.26 ±  20.10       & 0.60 ± 0.16           & 43.23 ± 14.22        & \textbf{1.00 ± 0.00}            \\
IQF         & 78.18 ± 2.43        & 0.85 ± 0.02             & 30.66 ± 22.62        & 0.73 ± 0.09            \\
IDDPG       &  -7.27 ± 12.41        & 0.39 ± 0.08            & 30.26 ± 6.83       & 0.79 ± 0.03            \\
MQF   & \textbf{94.61 ± 1.89}        & \textbf{0.95 ± 0.02}            & \textbf{91.31 ± 5.80 }       & 0.96 ± 0.02            \\
MADDPG-Ind  & -54.57 ± 18.05        & 0.195 ± 0.09            & -69.84 ± 13.37        & 0.13 ± 0.09            \\
MADDPG-Team & -20.78 ± 18.71        & 0.339 ± 0.12            & 10.03 ± 13.35        & 0.80 ± 0.08            
\end{tblr}
}

\caption{\small{Average metrics with 95\% confidence intervals for multiple runs on MWE upon training completion.}}
\label{table:experiment_results_walker}
\end{table}

Finally, we evaluate the algorithms' performance in MWE variations, specifically 2A1W and 4A2W configurations. 
The trained models exhibit multiple behavioral patterns, varying in their degree of optimality: (a) learning merely to hold the package without falling, (b) focusing on throwing the package as far as possible, (c) mastering balance and forward movement with small steps, and (d) successfully carrying the package to the terrain's end before the episode concludes. Behaviors (a) to (c) are indicative of the algorithms being stuck in suboptimal policies. Analysis of early-trained models reveals an initial tendency towards behavior (b), which gradually transitions to behaviors (a) and (c) as training progressed.

While independent learning algorithms, IQF and IDDPG, initially seem comparable to MQF in Fig.~\ref{fig:body_multi-walkers}, their limitations become apparent upon deeper analysis. These models exhibit a notable drop in average reward and stability with an increasing number of test episodes. In particular, although these algorithms, unlike MQF, rarely achieve the optimal behavior (d), their performance lack consistency, as shown by the average team rewards and balance rates (indicator of avoiding falls or package drops) presented in Table~\ref{table:experiment_results_walker}. CDDPG, however, despite recording a high balance rate in the 4A2W scenario, often fails to transport the package within the episode's maximum steps. In contrast, MQF consistently delivers higher returns in both scenarios. It is worth noting that extending training time or increasing the maximum step count per episode could potentially allow other methods to reach optimal solutions, a hypothesis supported by MQF's enhanced sample efficiency and overall effectiveness.

\section{Conclusion and Future Work}

We introduced a novel algorithm that leverages the strengths of value-based strategies to address sample inefficiency in policy-based methods, particularly in the context of cooperative multi-agent tasks within continuous action spaces. Our approach enables agents to characterize each state using a basis function, and optimize its coefficients. To further facilitate cooperation among agents, we employ a mixer network that combines their action-values effectively. Our empirical findings are twofold: (i) our method consistently achieves optimal solutions, surpassing policy-based methods which frequently converge to suboptimal cases; (ii) in instances where policy-based methods also attain optimal solutions, our method demonstrates a faster rate of convergence.

Looking ahead, our primary objective is to enhance the stability of the agents' learning model. Although our method already exhibits greater stability compared to existing approaches, we observed occasional fluctuations in some trials, suggesting potential instability. To address these challenges, future work can consider incorporating additional stability measures, such as target policy smoothing in~\cite{fujimoto2018addressing}, and adapting more sophisticated mixer networks, especially as the complexity of the environments escalates.

% \addtolength{\textheight}{-12.5cm}
\bibliographystyle{named}
\bibliography{ijcai24}

\clearpage
\appendix

\section{Supplementary Material}
\subsection{MQF Algorithm}
\label{MQF}

% \renewcommand{\algorithmicrequire}{\textbf{Input:}}
% \renewcommand{\algorithmicensure}{\textbf{Output:}}

% \SetAlCapNameFnt{\scriptsize}
% \SetAlCapFnt{\scriptsize}

% \begin{algorithm}[!h]
%     \scriptsize
%     \caption{Action Representations}
%     \label{alg:MQF}
%     \SetAlgoLined
%     \SetKwInOut{Input}{input}
%     \SetKwInOut{Output}{output}
%     \DontPrintSemicolon
    
%     \Input{order, $o$; actions, $a$}
%     $V$ $\leftarrow$ empty list\Comment{action representations}\;
%     $p$ $\leftarrow$ range($0$, $o$) \Comment{power list}\;
%     $d$ $\leftarrow$ action dimension\;
%     $p_\textup{combs}$ $\leftarrow$ cartesian product of $p$ and $d$ \Comment{all combinations of powers}\;
%     \vspace{0.05cm}
%     \ForEach{$p_\textup{comb}$ \textup{\textbf{in}} $p_\textup{combs}$}{
%     \vspace{0.05cm}
%     \If{sum($p_\textup{comb}$)$\leq$ o}{
%         \vspace{0.05cm}
%         $v$ $\leftarrow$ $1$ \Comment{initializing representation}\;
%         \vspace{0.05cm}
%         \For{$i=1, \ldots, d$}{
%             \vspace{0.05cm}
%             $v$ $\leftarrow$ $v * a[i]^{p_\textup{comb}[i]}$\;
%             \vspace{0.05cm}
%             }
%             \vspace{0.05cm}
%         Add $v$ to $V$ \Comment{adding the representation to the list of those}\;
%         }

%     }
% return $V$ \Comment{returning action representations}
% \end{algorithm}

\SetAlCapNameFnt{\scriptsize}
\SetAlCapFnt{\scriptsize}

\renewcommand{\algorithmicrequire}{\textbf{Input:}}
\renewcommand{\algorithmicensure}{\textbf{Output:}}
\SetAlCapNameFnt{\scriptsize}
\SetAlCapFnt{\scriptsize}

\begin{algorithm}[!h]
    \scriptsize
    % \caption{\small{Relational Q-Functionals}}
    \caption{Mixed Q-Functionals}
    \label{alg:MQF}
    \SetAlgoLined
    \SetKwInOut{Input}{input}
    \SetKwInOut{Output}{output}
    \DontPrintSemicolon
    
    \Input{prediction network, $\hat{Q}^{\textrm{F}}_{\textrm{prediction}}$; target network, $\hat{Q}^{\textrm{F}}_{\textrm{target}}$; relational graph, $G$; batch size, $b$; update frequency, \textit{step}$_\textrm{update}$; action range, $[\textit{a}_\textrm{min}, \textit{a}_\textrm{max}]$; action representation calculator, $\Phi$; order of the basis function, $o$;}
    \vspace{0.05cm}
    \ForEach{episode}{
        \vspace{0.05cm}
        Initialize s\Comment{$s \in \mathbb{R}^N$}\;
        \vspace{0.05cm}
        \ForEach{step \textup{\textbf{of}} episode}{
            
            % rand\_actions $\leftarrow$ $\mathcal{U}$($N$, act\_dim)\Comment{agents' actions}\;
            $a_\textrm{rand}$ $\leftarrow$ $\mathcal{U}$([$\textit{a}_\textrm{min}$, $\textit{a}_\textrm{max}$])\Comment{random action selection}\;
            \vspace{0.05cm}
            
            % coefficients $\leftarrow \hat{Q}^{\textrm{F}}_{\textrm{prediction}}$(states)\Comment{states' coefficients}\;
            
            % $f_\textrm{c}$ $\leftarrow \hat{Q}^{\textrm{F}}_{\textrm{prediction}}$(s)\Comment{ state function's coefficients}\;

            $C_s$ $\leftarrow \hat{Q}^{\textrm{F}}_{\textrm{prediction}}$(s)\Comment{state function's coefficients}\;

            \vspace{0.05cm}
            % actions\_repr $\leftarrow$ get\_repr(basis, order, rand\_actions)\Comment{representation}\;

            $V_s$ $\leftarrow$ $\Phi$($o$, $a_\textrm{rand}$)\Comment{action representations}\;
            %$f_\textrm{r}$ $\leftarrow$ get\_repr($f_\textrm{basis}$, $o$, $a_\textrm{random}$)\Comment{state function's representation}\;
            
            \vspace{0.05cm}
            
            % actions\_val $\leftarrow$ mat\_mult(coefficients, actions\_repr)\Comment{actions' value}\;
            $Q$ $\leftarrow$ $C_s V_s$\Comment{calculate action values using matrix multiplication}\;
            
            \vspace{0.05cm}
            $a_\textrm{best}$ $\leftarrow$ $\arg \max$($Q$)\Comment{best action for each agent}\;

            % \vspace{0.05cm}
            % actions\_noise $\leftarrow$ gaussian\_noise($N$, act\_dim)\Comment{actions' noise}\;
            
            % \vspace{0.05cm}
            % $a$ $\leftarrow$ best\_actions $ + $ actions\_noise\Comment{calculated actions}\;
            
            \vspace{0.05cm}
            $a$ $\leftarrow$  $a_{\textrm{best}} + \epsilon$ \Comment{$\epsilon \sim \mathcal{N}(0, 0.1)$}\;
            
            \vspace{0.05cm}
            Take actions, $a$, observe $r$, $s'$\;
            
            \vspace{0.05cm}
            Store $s$, $a$, $r$, $s'$ in memory\;
            
            \vspace{0.05cm}
            % Update $s$ with $s'$\;
            $s$ $\leftarrow$  $s'$\;
            
            \vspace{0.05cm}
            \textbf{if} mod(\textit{step}, \textit{step}$_\textrm{update}$) is $0$ \textbf{then} \Comment{network update process}
                        
            % \If{step \textup{\%} update\_step \textup{is zero}}{
                        \vspace{0.05cm}
                        \hspace{\algorithmicindent}$S$, $A$, $R$, $S'$ $\leftarrow$ sample chunk, size of $b$, from memory\;
                        
                        \vspace{0.02cm}
                        \hspace{\algorithmicindent}$C_s$ $\leftarrow \hat{Q}^{\textrm{F}}_{\textrm{prediction}}$($S$)\;
                        
                        \vspace{0.05cm}
                        \hspace{\algorithmicindent}$V_s$ $\leftarrow$ $\Phi$($o$, $A$)\;
                        
                        \vspace{0.05cm}
                        \hspace{\algorithmicindent}$Q^\textrm{prediction}\leftarrow$ $C_s V_s$\;
                        
                        \vspace{0.05cm}
                        \hspace{\algorithmicindent}$Q^\textrm{prediction}_{\textrm{team}} \leftarrow$ mix($Q^\textrm{prediction}$) \Comment{mixing action values}\;
                        
                        \vspace{0.05cm}
                        \hspace{\algorithmicindent}$a_\textrm{rand}$ $\leftarrow$ $\mathcal{U}$([$\textit{a}_\textrm{min}$, $\textit{a}_\textrm{max}$])\;
                        
                        \vspace{0.05cm}
                        \hspace{\algorithmicindent}$C_s$ $\leftarrow \hat{Q}^{\textrm{F}}_{\textrm{target}}$($S'$)\;
                        
                        \vspace{0.05cm}
                        \hspace{\algorithmicindent}$V_s$ $\leftarrow$ $\Phi$($o$, $a_\textrm{rand}$)\;
                        
                        \vspace{0.05cm}
                        \hspace{\algorithmicindent}$Q^\textrm{target}$ $\leftarrow$ $C_s V_s$\;
                        
                        \vspace{0.05cm}
                        \hspace{\algorithmicindent}$Q^\textrm{target}_\textrm{best} \leftarrow$ $\arg \max$($Q^\textrm{target}$)\;
                        
                        \vspace{0.1cm}
                        \hspace{\algorithmicindent}$Q^\textrm{target}_{\textrm{team}} \leftarrow $ mix($Q^\textrm{prediction}$)\Comment{mixing action values}\;
                        
                        \vspace{0.1cm}
                        \hspace{\algorithmicindent}$L \leftarrow$ use (9) with $R$, $Q^{\textrm{target}}_{\textrm{team}}$, $Q^{\textrm{prediction}}_{\textrm{team}}$\;
                        
                        \vspace{0.05cm}
                        \hspace{\algorithmicindent}Backpropagate $L$ to the parameters of $\hat{Q}^{\textrm{prediction}}$\;
                        
                        \vspace{0.05cm}
                        \hspace{\algorithmicindent}Update weights of $\hat{Q}^{\textrm{F}}_{\textrm{target}}$ with those of $\hat{Q}^{\textrm{F}}_{\textrm{prediction}}$\;
                        \vspace{0.05cm}

        }        
    }

\end{algorithm}

\newpage

\subsection{Environments}
\label{environments}

\begin{figure}[!h]
\centering
  \includegraphics[width=\linewidth]{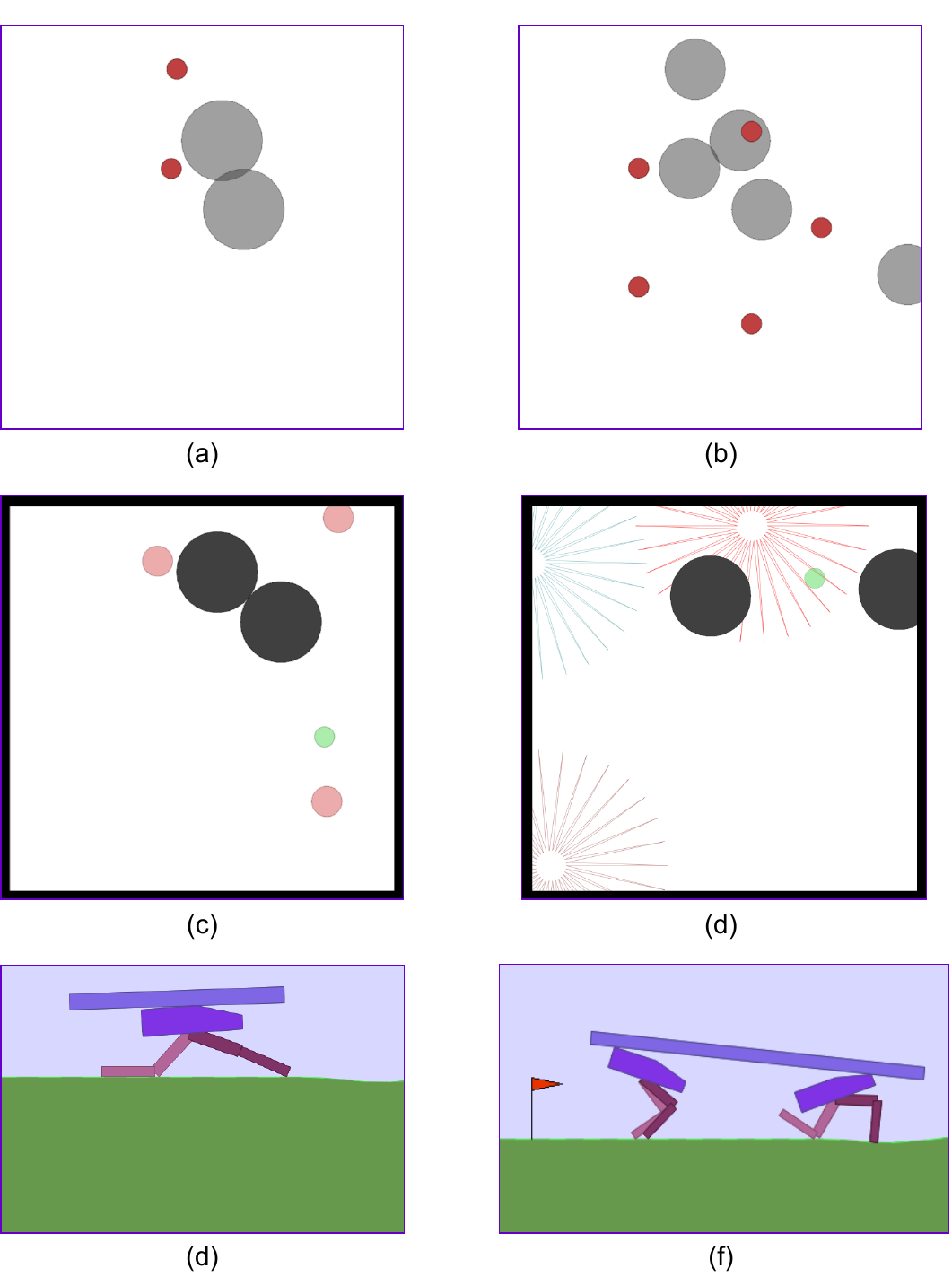}
  \caption{\small{Environments: (a) 2A2L, (b) 5A5L, (c) S-PP, (d) IC-PP, (e) 2A1W, and (f) 4A2W.}} 
    \label{fig:environments_image}
\end{figure}

\newpage

\subsection{Experimental Details}
\label{experimental_details}

\subsubsection{2 Agents \& 2 Landmarks (2A2L)}

% Environmental hyperparameters for 2A2L

\begin{table}[h]
\centering
\resizebox{\linewidth}{!}{
\begin{tblr}{
  row{3} = {c},
  row{4} = {c},
  row{5} = {c},
  row{6} = {c},
  row{7} = {c},
  row{8} = {c},
  row{9} = {c},
  row{10} = {c},
  row{11} = {c},
  row{12} = {c},
  row{13} = {c},
  row{14} = {c},
  row{15} = {c},
  row{16} = {c},
  row{17} = {c},
  row{18} = {c},
  row{19} = {c},
  row{20} = {c},
  cell{1}{2} = {c=7}{c},
  cell{2}{2} = {c},
  cell{2}{3} = {c},
  cell{2}{4} = {c},
  cell{2}{5} = {c},
  cell{2}{6} = {c},
  cell{2}{7} = {c},
  cell{2}{8} = {c},
  cell{20}{2} = {c},
  cell{20}{3} = {c},
  cell{20}{4} = {c},
  cell{20}{5} = {c},
  cell{20}{6} = {c},
  cell{20}{7} = {c},
  cell{20}{8} = {c},
  vline{2} = {1-20}{},
  hline{3} = {2-8}{},
}
                         \textbf{hyperparameters} & \textbf{Algorithms} &       &     &       &     &            &             \\
 & CQF                 & CDDPG & IQF & IDDPG & MQF & MADDPG-Ind & MADDPG-Team \\
\textit{max\_episode}             & 10000                   & 10000      & 10000    & 10000      & 10000    & 10000           & 10000            \\
\textit{nb\_runs}             & 5                   & 5      & 5    & 5      & 5    & 5           & 5            \\
\textit{gamma}                   & 0.99                    & 0.99      & 0.99    & 0.99      & 0.99    & 0.99           & 0.99            \\
\textit{exploration}              & e-greedy                    & ornstein-uhlenbeck      & e-greedy    & ornstein-uhlenbeck      & e-greedy    &   --         &  --           \\
\textit{learning\_starts\_steps}         & 10000                    & 10000      & 10000    & 10000      & 10000    & 10000           & 10000            \\
\textit{buffer\_size}             &  500000                   & 500000      & 500000    & 500000      & 500000    & 500000           & 500000            \\
\textit{sample\_method}           &   uniform                  & --      & uniform    & --      &uniform    & --           & --            \\
\textit{sample\_size}             &  1000                   & --      & 1000    & --      & 1000    & --           & --            \\
\textit{steps\_per\_update }      & 1                    & 10      & 1    & 50      & 1    &  50          & 50            \\
\textit{batch\_size}              & 512                    & 512      & 512    & 512      & 512    & 512           & 512            \\
\textit{optimizer}                & Adam                    & Adam      & Adam    & Adam      & Adam    & Adam           & Adam            \\
\textit{lr\_actor}                & 0.001                    &  0.000025     & 0.001    & 0.000025      & 0.001    & 0.001           & 0.001            \\
\textit{lr\_critic}               &  --                   & 0.00025      & --    & 0.00025      & --    & 0.001           & 0.001            \\
\textit{target\_network\_lr}      &  0.005                   & 0.005      & 0.005    & 0.005      & 0.005    &  0.01          & 0.01            \\
\textit{nb\_layers}               & 3                    & 2      & 3    & 2      & 3    & 3           & 3            \\
\textit{nb\_neurons }             & 256                    & 256      & 256    & 256      & 256    & 256           & 256            \\
\textit{activation\_actor}        & tanh                    & tanh      & tanh    & tanh      & tanh    & relu           & relu            \\
\textit{activation\_critic}       & --                    & relu      & --    & relu      & --    & relu           & relu            \\

                         &                     &       &     &       &     &            &             \\
                         &                     &       &     &       &     &            &             
\end{tblr}
}
\caption{\small{Algorithmic hyperparameters for 2A2L.}}
\label{table:appendix_2A2L}
\end{table}

% \vspace{0.15cm}

\begin{figure}[h]
\centering
  \includegraphics[width=\linewidth]{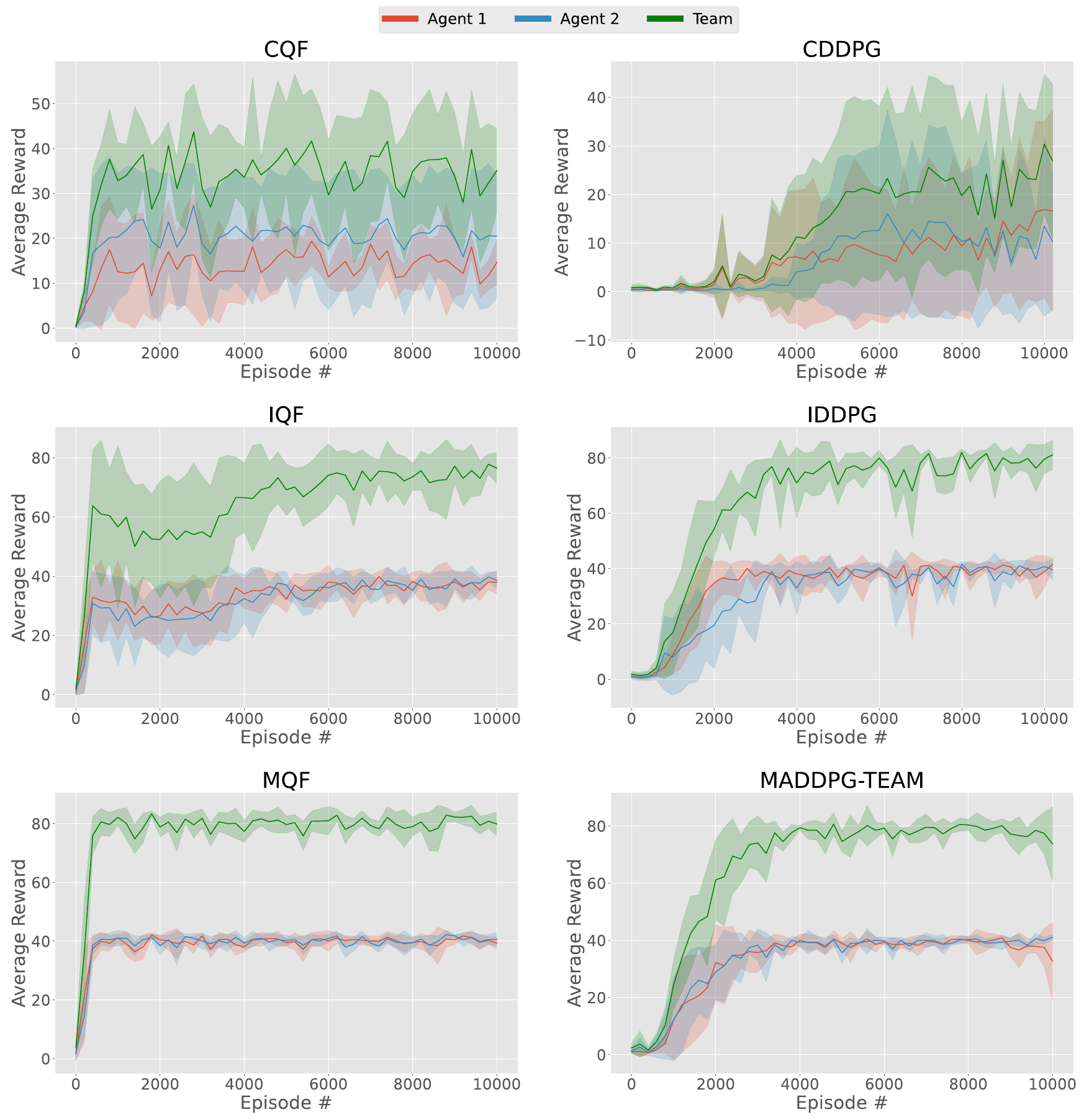}
  \caption{\small{Detailed results for 2A2L across CQF, CDDPG, IQF, IDDPG, MQF, and MADDPG-Team.}} 
    \label{fig:appendix_2A2L}
\end{figure}

\newpage

\vspace*{0.0001\baselineskip}
\subsubsection{5 Agents \& 5 Landmarks (5A5L)}

% Environmental hyperparameters for 5A5L

\begin{table}[!h]
\centering
\resizebox{\linewidth}{!}{
\begin{tblr}{
  row{3} = {c},
  row{4} = {c},
  row{5} = {c},
  row{6} = {c},
  row{7} = {c},
  row{8} = {c},
  row{9} = {c},
  row{10} = {c},
  row{11} = {c},
  row{12} = {c},
  row{13} = {c},
  row{14} = {c},
  row{15} = {c},
  row{16} = {c},
  row{17} = {c},
  row{18} = {c},
  row{19} = {c},
  row{20} = {c},
  cell{1}{2} = {c=7}{c},
  cell{2}{2} = {c},
  cell{2}{3} = {c},
  cell{2}{4} = {c},
  cell{2}{5} = {c},
  cell{2}{6} = {c},
  cell{2}{7} = {c},
  cell{2}{8} = {c},
  cell{20}{2} = {c},
  cell{20}{3} = {c},
  cell{20}{4} = {c},
  cell{20}{5} = {c},
  cell{20}{6} = {c},
  cell{20}{7} = {c},
  cell{20}{8} = {c},
  vline{2} = {1-20}{},
  hline{3} = {2-8}{},
}
                         \textbf{hyperparameters} & \textbf{Algorithms} &       &     &       &     &            &             \\
 & CQF                 & CDDPG & IQF & IDDPG & MQF & MADDPG-Ind & MADDPG-Team \\
\textit{max\_episode}             & 10000                   & 10000      & 10000    & 10000      & 10000    & 10000           & 10000            \\
\textit{nb\_runs}             & 5                   & 5      & 5    & 5      & 5    & 5           & 5            \\
\textit{gamma}                   & 0.99                    & 0.99      & 0.99    & 0.99      & 0.99    & 0.99           & 0.99            \\
\textit{exploration}              & e-greedy                    & ornstein-uhlenbeck      & e-greedy    & ornstein-uhlenbeck      & e-greedy    &   --         &  --           \\
\textit{learning\_starts\_steps}         & 10000                    & 10000      & 10000    & 10000      & 10000    & 10000           & 10000            \\
\textit{buffer\_size}             &  500000                   & 500000      & 500000    & 500000      & 500000    & 500000           & 500000            \\
\textit{sample\_method}           &   uniform                  & --      & uniform    & --      &uniform    & --           & --            \\
\textit{sample\_size}             &  1000                   & --      & 1000    & --      & 1000    & --           & --            \\
\textit{steps\_per\_update }      & 1                    & 10      & 1    & 50      & 1    &  50          & 50            \\
\textit{batch\_size}              & 512                    & 512      & 512    & 512      & 512    & 512           & 512            \\
\textit{optimizer}                & Adam                    & Adam      & Adam    & Adam      & Adam    & Adam           & Adam            \\
\textit{lr\_actor}                & 0.001                    &  0.000025     & 0.001    & 0.000025      & 0.001    & 0.001           & 0.001            \\
\textit{lr\_critic}               &  --                   & 0.00025      & --    & 0.00025      & --    & 0.001           & 0.001            \\
\textit{target\_network\_lr}      &  0.005                   & 0.005      & 0.005    & 0.005      & 0.005    &  0.01          & 0.01            \\
\textit{nb\_layers}               & 3                    & 2      & 3    & 2      & 3    & 3           & 3            \\
\textit{nb\_neurons }             & 256                    & 256      & 256    & 256      & 256    & 256           & 256            \\
\textit{activation\_actor}        & tanh                    & tanh      & tanh    & tanh      & tanh    & relu           & relu            \\
\textit{activation\_critic}       & --                    & relu      & --    & relu      & --    & relu           & relu            \\

                         &                     &       &     &       &     &            &             \\
                         &                     &       &     &       &     &            &             
\end{tblr}
}
\label{table:appendix_5A5L}
\caption{\small{Algorithmic hyperparameters for 5A5L.}}
\end{table}

\begin{figure}[!h]
\centering
  \includegraphics[width=\linewidth]{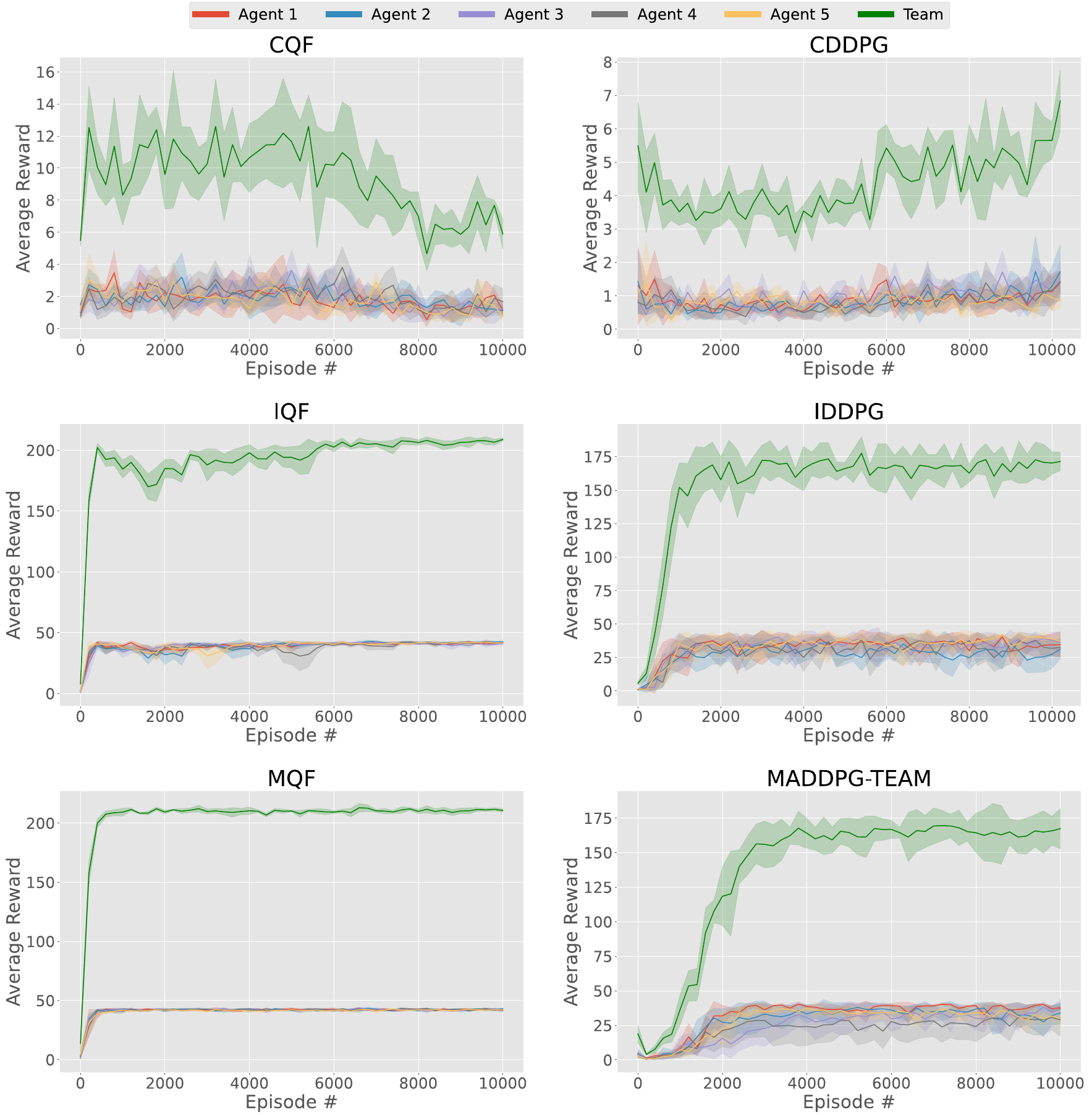}
  \caption{\small{Detailed results for 5A5L across CQF, CDDPG, IQF, IDDPG, MQF, and MADDPG-Team.}} 
    \label{fig:appendix_5A5L}
\end{figure}

\newpage

\subsubsection{Standard Predator-Prey (S-PP)}

% Environmental hyperparameters for S-PP

\begin{table}[!h]
\centering
\resizebox{\linewidth}{!}{
\begin{tblr}{
  row{3} = {c},
  row{4} = {c},
  row{5} = {c},
  row{6} = {c},
  row{7} = {c},
  row{8} = {c},
  row{9} = {c},
  row{10} = {c},
  row{11} = {c},
  row{12} = {c},
  row{13} = {c},
  row{14} = {c},
  row{15} = {c},
  row{16} = {c},
  row{17} = {c},
  row{18} = {c},
  row{19} = {c},
  row{20} = {c},
  cell{1}{2} = {c=7}{c},
  cell{2}{2} = {c},
  cell{2}{3} = {c},
  cell{2}{4} = {c},
  cell{2}{5} = {c},
  cell{2}{6} = {c},
  cell{2}{7} = {c},
  cell{2}{8} = {c},
  cell{20}{2} = {c},
  cell{20}{3} = {c},
  cell{20}{4} = {c},
  cell{20}{5} = {c},
  cell{20}{6} = {c},
  cell{20}{7} = {c},
  cell{20}{8} = {c},
  vline{2} = {1-20}{},
  hline{3} = {2-8}{},
}
                         \textbf{hyperparameters} & \textbf{Algorithms} &       &     &       &     &            &             \\
 & CQF                 & CDDPG & IQF & IDDPG & MQF & MADDPG-Ind & MADDPG-Team \\
\textit{max\_episode}             & 120000                   & 120000      & 120000    & 120000      & 120000    & 120000           & 120000            \\
\textit{nb\_runs}             & 3                   & 3      & 3    & 3      & 3    & 3           & 3            \\
\textit{gamma}                   & 0.99                    & 0.99      & 0.99    & 0.99      & 0.99    & 0.99           & 0.99            \\
\textit{exploration}              & gaussian                    & ornstein-uhlenbeck      & gaussian    & ornstein-uhlenbeck      & gaussian    &   --         &  --           \\
\textit{learning\_starts\_steps}         & 10000                    & 10000      & 10000    & 10000      & 10000    & 10000           & 10000            \\
\textit{buffer\_size}             &  500000                   & 500000      & 500000    & 500000      & 500000    & 500000           & 500000            \\
\textit{sample\_method}           &   uniform                  & --      & uniform    & --      &uniform    & --           & --            \\
\textit{sample\_size}             &  1000                   & --      & 1000    & --      & 1000    & --           & --            \\
\textit{steps\_per\_update }      & 50                    & 10      & 50    & 50      & 50    &  50          & 50            \\
\textit{batch\_size}              & 512                    & 512      & 512    & 512      & 512    & 512           & 512            \\
\textit{optimizer}                & Adam                    & Adam      & Adam    & Adam      & Adam    & Adam           & Adam            \\
\textit{lr\_actor}                & 0.001                    &  0.000025     & 0.001    & 0.000025      & 0.001    & 0.001           & 0.001            \\
\textit{lr\_critic}               &  --                   & 0.00025      & --    & 0.00025      & --    & 0.001           & 0.001            \\
\textit{target\_network\_lr}      &  0.005                   & 0.005      & 0.005    & 0.005      & 0.005    &  0.01          & 0.01            \\
\textit{nb\_layers}               & 3                    & 2      & 3    & 2      & 3    & 3           & 3            \\
\textit{nb\_neurons }             & 256                    & 256      & 256    & 256      & 256    & 256           & 256            \\
\textit{activation\_actor}        & tanh                    & tanh      & tanh    & tanh      & tanh    & relu           & relu            \\
\textit{activation\_critic}       & --                    & relu      & --    & relu      & --    & relu           & relu            \\

                         &                     &       &     &       &     &            &             \\
                         &                     &       &     &       &     &            &             
\end{tblr}
}
\caption{\small{Algorithmic hyperparameters for S-PP.}}
\label{experiment_results_walker}
\end{table}

\begin{figure}[!h]
\centering
  \includegraphics[width=\linewidth]{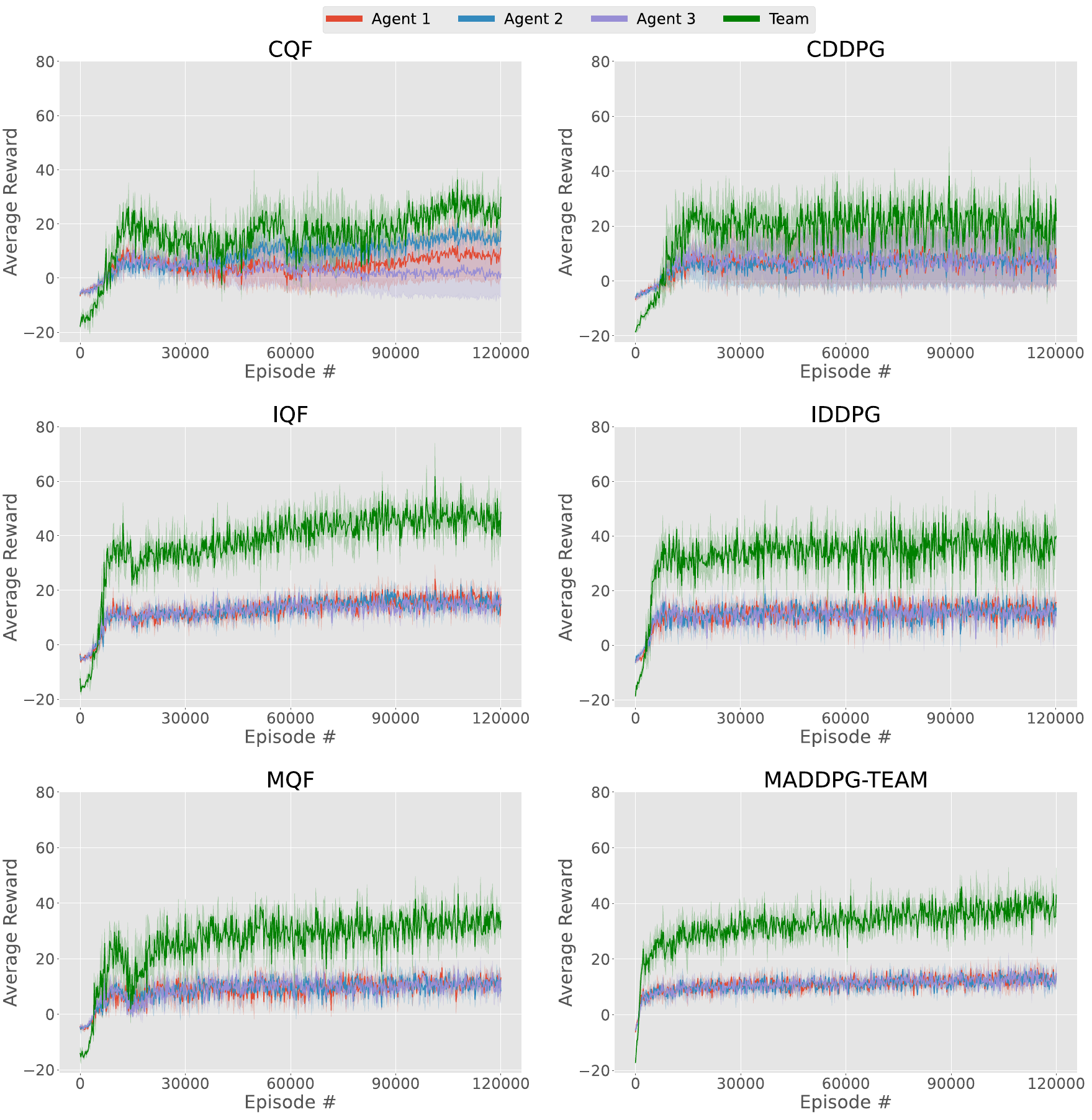}
  \caption{\small{Detailed results for S-PP across CQF, CDDPG, IQF, IDDPG, MQF, MADDPG-Team.}} 
  % with agents' individual reward
    \label{fig:appendix_S-PP}
\end{figure}

\newpage

\subsubsection{Increased Cooperation Predators-Prey (IC-PP)}

% Environmental hyperparameters for IC-PP

\begin{table}[!h]
\centering
\resizebox{\linewidth}{!}{
\begin{tblr}{
  row{3} = {c},
  row{4} = {c},
  row{5} = {c},
  row{6} = {c},
  row{7} = {c},
  row{8} = {c},
  row{9} = {c},
  row{10} = {c},
  row{11} = {c},
  row{12} = {c},
  row{13} = {c},
  row{14} = {c},
  row{15} = {c},
  row{16} = {c},
  row{17} = {c},
  row{18} = {c},
  row{19} = {c},
  row{20} = {c},
  cell{1}{2} = {c=7}{c},
  cell{2}{2} = {c},
  cell{2}{3} = {c},
  cell{2}{4} = {c},
  cell{2}{5} = {c},
  cell{2}{6} = {c},
  cell{2}{7} = {c},
  cell{2}{8} = {c},
  cell{20}{2} = {c},
  cell{20}{3} = {c},
  cell{20}{4} = {c},
  cell{20}{5} = {c},
  cell{20}{6} = {c},
  cell{20}{7} = {c},
  cell{20}{8} = {c},
  vline{2} = {1-20}{},
  hline{3} = {2-8}{},
}
                         \textbf{hyperparameters} & \textbf{Algorithms} &       &     &       &     &            &             \\
 & CQF                 & CDDPG & IQF & IDDPG & MQF & MADDPG-Ind & MADDPG-Team \\
\textit{max\_episode}             & 40000                   & 40000      & 40000    & 40000      & 40000    & 40000           & 40000            \\
\textit{nb\_runs}             & 3                   & 3      & 3    & 3      & 3    & 3           & 3            \\
\textit{gamma}                   & 0.99                    & 0.99      & 0.99    & 0.99      & 0.99    & 0.99           & 0.99            \\
\textit{exploration}              & gaussian                    & ornstein-uhlenbeck      & gaussian    & ornstein-uhlenbeck      & gaussian    &   --         &  --           \\
\textit{learning\_starts\_steps}         & 10000                    & 10000      & 10000    & 10000      & 10000    & 10000           & 10000            \\
\textit{buffer\_size}             &  500000                   & 500000      & 500000    & 500000      & 500000    & 500000           & 500000            \\
\textit{sample\_method}           &   uniform                  & --      & uniform    & --      &uniform    & --           & --            \\
\textit{sample\_size}             &  1000                   & --      & 1000    & --      & 1000    & --           & --            \\
\textit{steps\_per\_update }      & 50                    & 50      & 50    & 10      & 50    &  50          & 50            \\
\textit{batch\_size}              & 512                    & 512      & 512    & 512      & 512    & 512           & 512            \\
\textit{optimizer}                & Adam                    & Adam      & Adam    & Adam      & Adam    & Adam           & Adam            \\
\textit{lr\_actor}                & 0.001                    &  0.000025     & 0.001    & 0.000025      & 0.001    & 0.001           & 0.001            \\
\textit{lr\_critic}               &  --                   & 0.00025      & --    & 0.00025      & --    & 0.001           & 0.001            \\
\textit{target\_network\_lr}      &  0.005                   & 0.005      & 0.005    & 0.005      & 0.005    &  0.01          & 0.01            \\
\textit{nb\_layers}               & 3                    & 2      & 3    & 2      & 3    & 3           & 3            \\
\textit{nb\_neurons }             & 256                    & 256      & 256    & 256      & 256    & 256           & 256            \\
\textit{activation\_actor}        & tanh                    & tanh      & tanh    & tanh      & tanh    & relu           & relu            \\
\textit{activation\_critic}       & --                    & relu      & --    & relu      & --    & relu           & relu            \\

                         &                     &       &     &       &     &            &             \\
                         &                     &       &     &       &     &            &             
\end{tblr}
}
\label{table:appendix_S-PP}
\caption{\small{Algorithmic hyperparameters for IC-PP.}}
\end{table}

\begin{figure}[!h]
\centering
  \includegraphics[width=\linewidth]{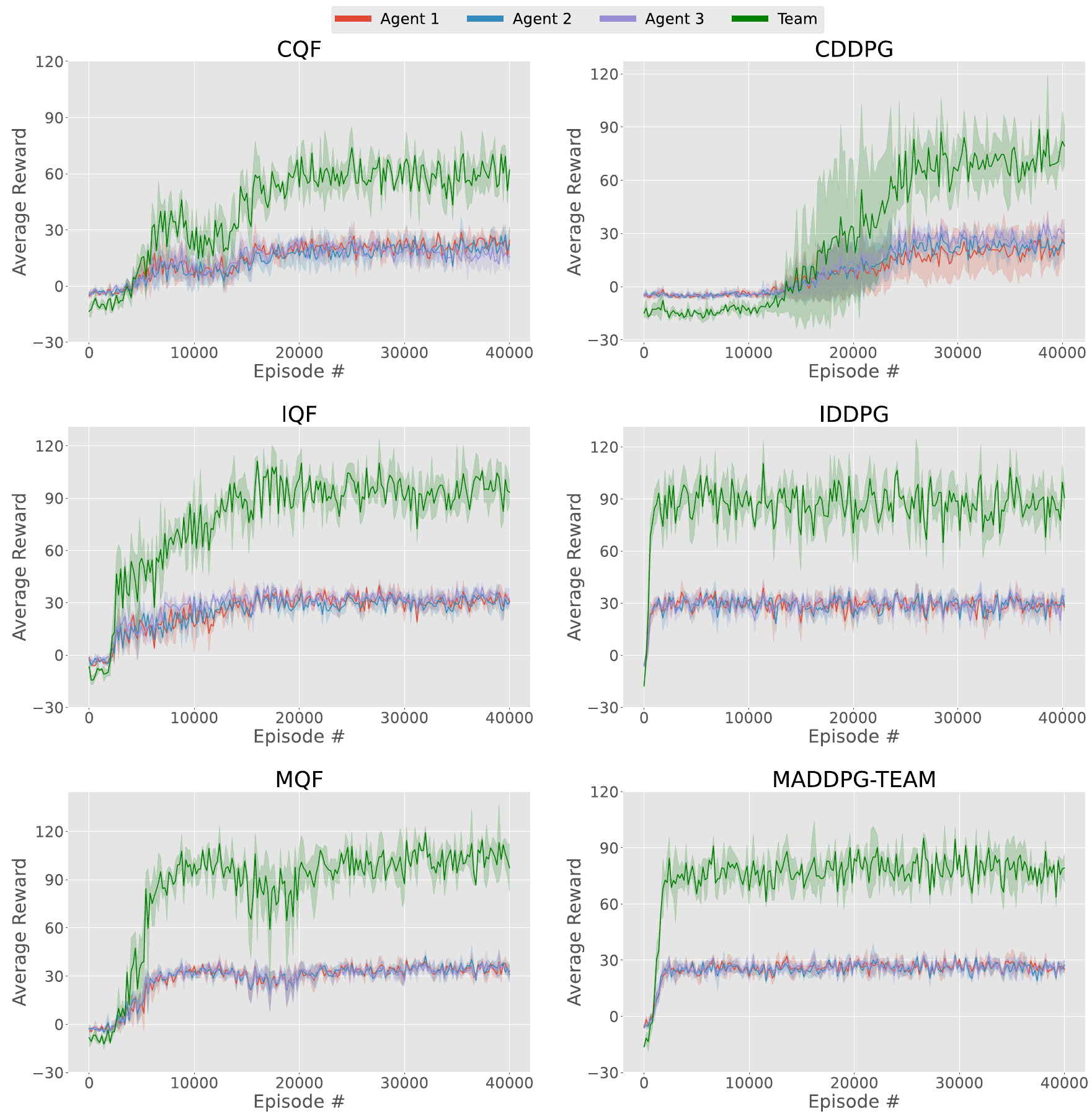}
  \caption{\small{Detailed results for S-PP across CQF, CDDPG, IQF, IDDPG, MQF, MADDPG-Team.}} 
  % with agents' individual reward
    \label{fig:appendix_IC-PP}
\end{figure}

\newpage

\subsubsection{Multi Walker Environment -- 2 Agents (2A1W)}

\begin{table}[!h]
\centering
\resizebox{0.35\linewidth}{!}{
\begin{tblr}{
    cells = {c},
    vline{2} = {-}{},
}
n\_walkers      & 1     \\
shared\_reward  & False \\
terrain\_length & 100   \\
max\_cycles     & 500   \\
position\_noise & 0     \\
angle\_noise    & 0     \\
forward\_reward & 5
\end{tblr}
}
\caption{\small{Environmental hyperparameters for 2A1W.}}
\end{table}
\begin{table}[!h]
\centering
\resizebox{\linewidth}{!}{
\begin{tblr}{
  row{3} = {c},
  row{4} = {c},
  row{5} = {c},
  row{6} = {c},
  row{7} = {c},
  row{8} = {c},
  row{9} = {c},
  row{10} = {c},
  row{11} = {c},
  row{12} = {c},
  row{13} = {c},
  row{14} = {c},
  row{15} = {c},
  row{16} = {c},
  row{17} = {c},
  row{18} = {c},
  row{19} = {c},
  row{20} = {c},
  cell{1}{2} = {c=7}{c},
  cell{2}{2} = {c},
  cell{2}{3} = {c},
  cell{2}{4} = {c},
  cell{2}{5} = {c},
  cell{2}{6} = {c},
  cell{2}{7} = {c},
  cell{2}{8} = {c},
  cell{20}{2} = {c},
  cell{20}{3} = {c},
  cell{20}{4} = {c},
  cell{20}{5} = {c},
  cell{20}{6} = {c},
  cell{20}{7} = {c},
  cell{20}{8} = {c},
  vline{2} = {1-20}{},
  hline{3} = {2-8}{},
}
                         \textbf{hyperparameters} & \textbf{Algorithms} &       &     &       &     &            &             \\
 & CQF                 & CDDPG & IQF & IDDPG & MQF & MADDPG-Ind & MADDPG-Team \\
\textit{max\_episode}             & 20000                   & 20000      & 20000    & 20000      & 20000    & 20000           & 20000            \\
\textit{nb\_runs}             & 10                   & 10      & 10    & 10      & 10    & 10           & 10            \\
\textit{gamma}                   & 0.99                    & 0.99      & 0.99    & 0.99      & 0.99    & 0.99           & 0.99            \\
\textit{exploration}              & gaussian                    & ornstein-uhlenbeck      & gaussian    & ornstein-uhlenbeck      & gaussian    &   --         &  --           \\
\textit{learning\_starts\_steps}         & 10000                    & 10000      & 10000    & 10000      & 10000    & 10000           & 10000            \\
\textit{buffer\_size}             &  500000                   & 500000      & 500000    & 500000      & 500000    & 500000           & 500000            \\
\textit{sample\_method}           &   uniform                  & --      & uniform    & --      &uniform    & --           & --            \\
\textit{sample\_size}             &  10000                   & --      & 10000    & --      & 10000    & --           & --            \\
\textit{steps\_per\_update }      & 50                    & 10      & 50    & 10      & 50    &  50          & 50            \\
\textit{batch\_size}              & 512                    & 512      & 512    & 512      & 512    & 512           & 512            \\
\textit{optimizer}                & Adam                    & Adam      & Adam    & Adam      & Adam    & Adam           & Adam            \\
\textit{lr\_actor}                & 0.001                    &  0.000025     & 0.001    & 0.000025      & 0.001    & 0.001           & 0.001            \\
\textit{lr\_critic}               &  --                   & 0.00025      & --    & 0.00025      & --    & 0.01           & 0.01            \\
\textit{target\_network\_lr}      &  0.01                   & 0.005      & 0.01    & 0.005      & 0.01    &  0.01          & 0.01            \\
\textit{nb\_layers}               & 2                    & 2      & 2    & 2      & 2    & 3           & 3            \\
\textit{nb\_neurons }             & 256                    & 256      & 256    & 256      & 256    & 256           & 256            \\
\textit{activation\_actor}        & tanh                    & tanh      & tanh    & tanh      & tanh    & tanh           & tanh            \\
\textit{activation\_critic}       & --                    & relu      & --    & relu      & --    & relu           & relu            \\

                         &                     &       &     &       &     &            &             \\
                         &                     &       &     &       &     &            &             
\end{tblr}
}
\label{table:appendix_S-PP}
\caption{\small{Algorithmic hyperparameters for 2A1W.}}
\end{table}
\begin{figure}[!h]
\centering
  \includegraphics[width=\linewidth]{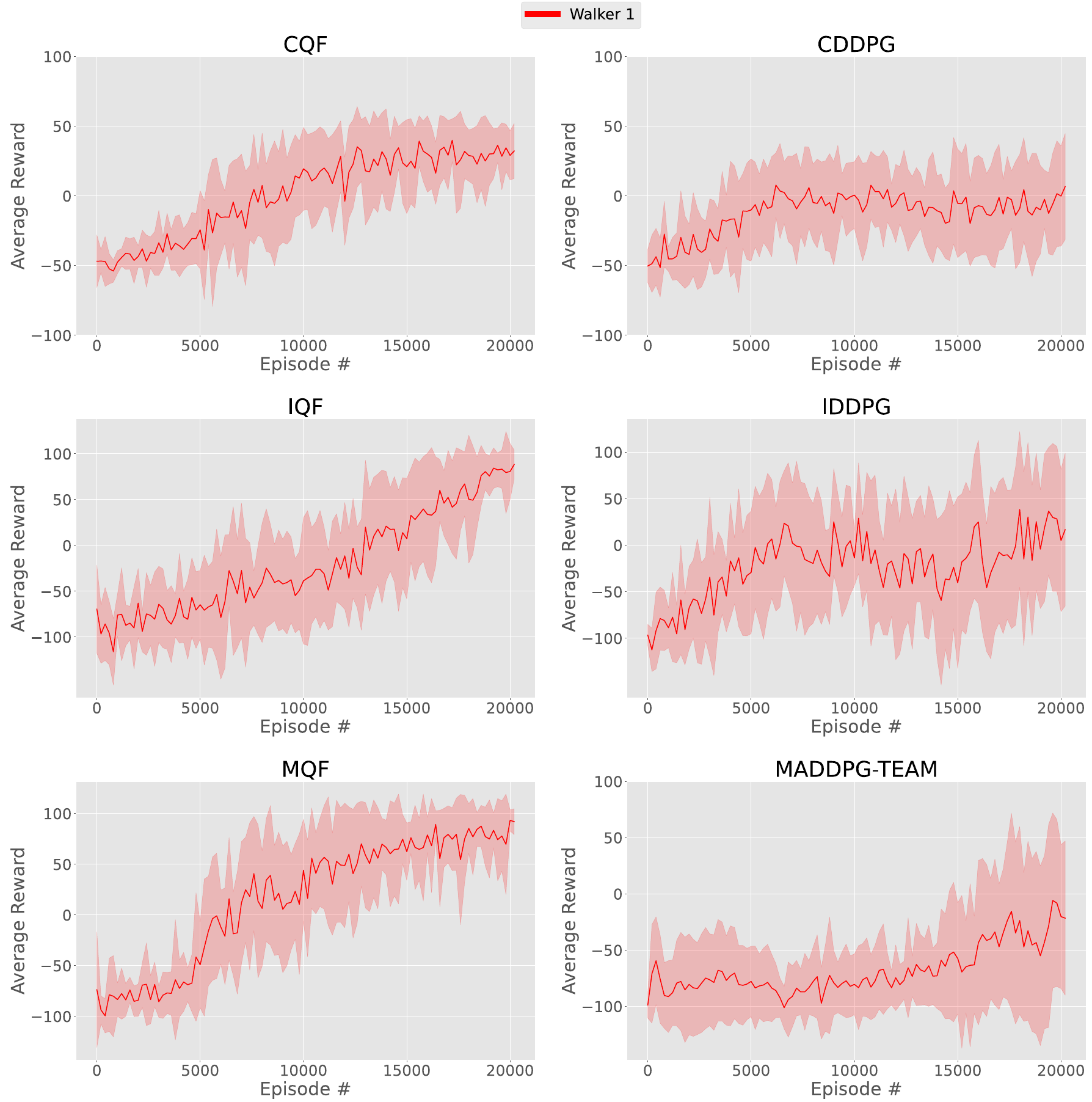}
  \caption{\small{Detailed results for 2A1W across CQF, CDDPG, IQF, IDDPG, MQF, MADDPG-Team. Since each joint (agent) receives the same reward as the corresponding walker, only the rewards for the walker is displayed.}} 
  % with agents' individual reward
    \label{fig:appendix_2A1W}
\end{figure}

\newpage

\subsubsection{Multi Walker Environment -- 4 Agents (4A2W)}

\begin{table}[!h]
\centering
\resizebox{0.35\linewidth}{!}{
\begin{tblr}{
  cells = {c},
    vline{2} = {-}{},
}
n\_walkers      & 2     \\
shared\_reward  & False \\
terrain\_length & 100   \\
max\_cycles     & 500   \\
position\_noise & 0     \\
angle\_noise    & 0     \\
forward\_reward & 5
\end{tblr}
}
\caption{\small{Environmental hyperparameters for 4A2W.}}
\end{table}
\begin{table}[!h]
\centering
\resizebox{\linewidth}{!}{
\begin{tblr}{
  row{3} = {c},
  row{4} = {c},
  row{5} = {c},
  row{6} = {c},
  row{7} = {c},
  row{8} = {c},
  row{9} = {c},
  row{10} = {c},
  row{11} = {c},
  row{12} = {c},
  row{13} = {c},
  row{14} = {c},
  row{15} = {c},
  row{16} = {c},
  row{17} = {c},
  row{18} = {c},
  row{19} = {c},
  row{20} = {c},
  cell{1}{2} = {c=7}{c},
  cell{2}{2} = {c},
  cell{2}{3} = {c},
  cell{2}{4} = {c},
  cell{2}{5} = {c},
  cell{2}{6} = {c},
  cell{2}{7} = {c},
  cell{2}{8} = {c},
  cell{20}{2} = {c},
  cell{20}{3} = {c},
  cell{20}{4} = {c},
  cell{20}{5} = {c},
  cell{20}{6} = {c},
  cell{20}{7} = {c},
  cell{20}{8} = {c},
  vline{2} = {1-20}{},
  hline{3} = {2-8}{},
}
                         \textbf{hyperparameters} & \textbf{Algorithms} &       &     &       &     &            &             \\
 & CQF                 & CDDPG & IQF & IDDPG & MQF & MADDPG-Ind & MADDPG-Team \\
\textit{max\_episode}             & 30000                   & 30000      & 30000    & 30000      & 30000    & 30000           & 30000            \\
\textit{nb\_runs}             & 5                   & 5      & 5    & 5      & 5    & 5           & 5            \\
\textit{gamma}                   & 0.99                    & 0.99      & 0.99    & 0.99      & 0.99    & 0.99           & 0.99            \\
\textit{exploration}              & gaussian                    & ornstein-uhlenbeck      & gaussian    & ornstein-uhlenbeck      & gaussian    &   --         &  --           \\
\textit{learning\_starts\_steps}         & 10000                    & 10000      & 10000    & 10000      & 10000    & 10000           & 10000            \\
\textit{buffer\_size}             &  500000                   & 500000      & 500000    & 500000      & 500000    & 500000           & 500000            \\
\textit{sample\_method}           &   uniform                  & --      & uniform    & --      &uniform    & --           & --            \\
\textit{sample\_size}             &  10000                   & --      & 10000    & --      & 10000    & --           & --            \\
\textit{steps\_per\_update }      & 50                    & 10      & 50    & 10      & 50    &  50          & 50            \\
\textit{batch\_size}              & 512                    & 512      & 512    & 512      & 512    & 512           & 512            \\
\textit{optimizer}                & Adam                    & Adam      & Adam    & Adam      & Adam    & Adam           & Adam            \\
\textit{lr\_actor}                & 0.001                    &  0.000025     & 0.001    & 0.000025      & 0.001    & 0.001           & 0.001            \\
\textit{lr\_critic}               &  --                   & 0.00025      & --    & 0.00025      & --    & 0.01           & 0.01            \\
\textit{target\_network\_lr}      &  0.01                   & 0.005      & 0.01    & 0.005      & 0.01    &  0.01          & 0.01            \\
\textit{nb\_layers}               & 2                    & 2      & 2    & 2      & 2    & 3           & 3            \\
\textit{nb\_neurons }             & 256                    & 256      & 256    & 256      & 256    & 256           & 256            \\
\textit{activation\_actor}        & tanh                    & tanh      & tanh    & tanh      & tanh    & tanh           & tanh            \\
\textit{activation\_critic}       & --                    & relu      & --    & relu      & --    & relu           & relu            \\

                         &                     &       &     &       &     &            &             \\
                         &                     &       &     &       &     &            &
\end{tblr}
}
\label{table:appendix_S-PP}
\caption{\small{Algorithmic hyperparameters for 4A2W.}}
\end{table}
\begin{figure}[!h]
\centering
  \includegraphics[width=\linewidth]{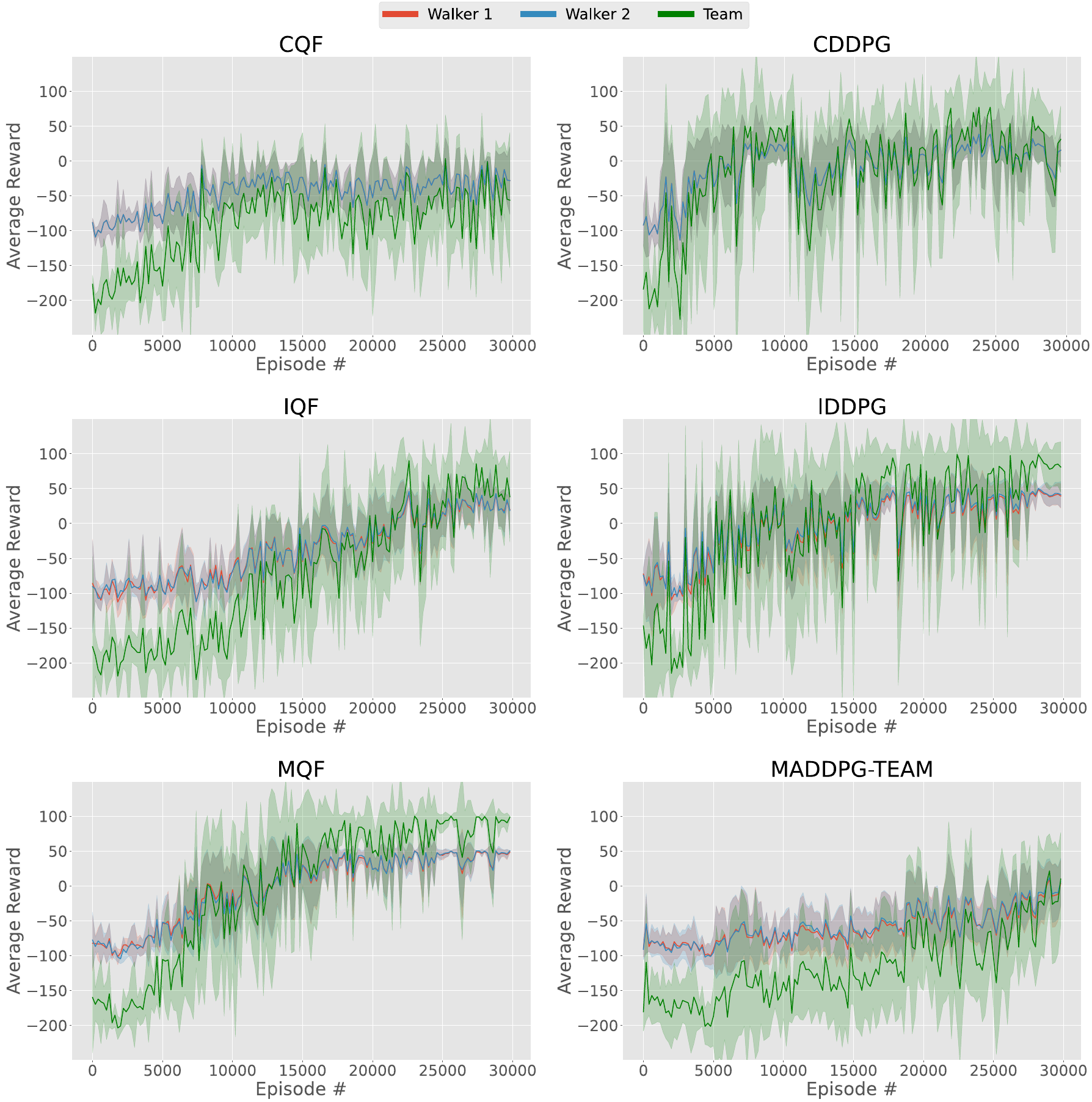}
  \caption{\small{Detailed results for 4A2W across CQF, CDDPG, IQF, IDDPG, MQF, MADDPG-Team. Since each joint (agent) receives the same reward as the corresponding walker, only the rewards for the walkers are displayed.}}
  % with agents' individual reward
    \label{fig:appendix_4A2W}
\end{figure}
\phantom{xxxx}
\let\clearpage\relax
\subsubsection{}

\end{document}